\documentclass[12pt,preprint]{aastex}
\pdfoutput=1

\usepackage{natbib}
\shorttitle{The Magnetic Field of NGC 1569}
\shortauthors{Kepley et al.}

\def\ha{H$\alpha \ $}
\def\kms{km~s$^{-1}\ $}

\newcommand{\uG}{\ensuremath{\rm{\mu  G \ }}}
\def\uJybeam{$\rm{\mu Jy \ beam^{-1}} \ $}
\def\mJybeam{$\rm{mJy \ beam^{-1}} \ $}

\begin{document}


\title{The Role of the Magnetic Field in the Interstellar Medium \\
  of the Post-Starburst Dwarf Irregular Galaxy NGC 1569}

\author{Amanda A. Kepley}
\affil{Department of Astronomy, University of Wisconsin--Madison, 475 North Charter Street, Madison, WI 53706} 
\email{kepley@astro.wisc.edu}

\author{Stefanie M\"uhle}
\affil{Joint Institute for VLBI in Europe, Postbus 2, 7990 AA Dwingeloo, The Netherlands}
\email{muehle@jive.nl}

\author{John Everett}
\affil{Department of Astronomy, University of Wisconsin--Madison, 475 North Charter Street, Madison, WI 53706}
\email{everett@physics.wisc.edu}

\author{Ellen G. Zweibel}
\affil{Department of Astronomy, University of Wisconsin--Madison, 475 North Charter Street, Madison, WI 53706}
\email{zweibel@astro.wisc.edu}

\author{Eric M. Wilcots}
\affil{Department of Astronomy, University of Wisconsin--Madison, 475 North Charter Street, Madison, WI 53706}
\email{ewilcots@astro.wisc.edu}

\author{Uli Klein}
\affil{Argelander-Institut f\"ur Astronomie, Universit\"at Bonn, Auf dem H\"ugel 71, D-53121 Bonn, Germany}
\email{uklein@astro.uni-bonn.de}

\begin{abstract}

  NGC 1569 is a nearby dwarf irregular galaxy which underwent an
  intense burst of star formation 10 to 40 Myr ago.  We present
  observations that reach surface brightnesses two to eighty times
  fainter than previous radio continuum observations and the first
  radio continuum polarization observations of this galaxy at 20\,cm,
  13\,cm, 6\,cm, and 3\,cm. These observations allow us to probe the
  relationship of the magnetic field of NGC 1569 to the rest of its
  interstellar medium. We confirm the presence of an extended radio
  continuum halo at 20\,cm and see for the first time the radio
  continuum feature associated with the western \ha arm at wavelengths
  shorter than 20\,cm. Although, in general, the spectral indices
  derived for this galaxy steepen as one moves into the halo of the
  galaxy, there are filamentary regions of flat spectral indices
  extending to the edge of the galaxy. The spectral index trends in
  this galaxy support the theory that there is a convective wind at
  work in this galaxy.  There is strong polarized emission at 3\,cm
  and 6\,cm and weak polarized emission at 20\,cm and 13\,cm.  We
  estimate that the thermal fraction is 40--50\% in the center of the
  galaxy and falls off rapidly with height above the disk. Using this
  estimate, we derive a total magnetic field strength of 38~\uG in the
  central regions and 10--15~\uG in the halo. The magnetic field is
  largely random in the center of the galaxy; the uniform field is
  $\sim$~3--9~\uG and is strongest in the halo. Using our total
  magnetic field strength estimates and the results of previous
  observations of NGC 1569, we find that the magnetic pressure is the
  same order of magnitude but, in general, a factor of a few less than
  the other components of the interstellar medium in this galaxy. The
  uniform magnetic field in NGC 1569 is closely associated with the
  \ha bubbles and filaments. We suggest that a supernova-driven dynamo
  may be operating in this galaxy. Based on our pressure estimates and
  the morphology of the magnetic field, the outflow of hot gas from
  NGC 1569 is clearly shaping the magnetic field, but the magnetic
  field in turn may be aiding the outflow by channeling gas out of the
  disk of the galaxy. Dwarf galaxies with extended radio continuum
  halos like that of NGC 1569 may play an important role in
  magnetizing the intergalactic medium.

\end{abstract}

\keywords{ galaxies: individual (NGC 1569) --- galaxies: ISM ---
galaxies: irregular --- galaxies: magnetic fields --- galaxies:
starburst --- radio continuum: galaxies }


\section{Introduction}

The structure of a galaxy's interstellar medium (ISM) is shaped by its
stars. Stars dissociate and ionize the surrounding cocoon of cold gas
from which they formed. Winds from massive stars and supernovae
explosions also create bubbles of hot gas.  The combined effects of
multiple stars in a cluster create larger bubbles of hot gas referred
to superbubbles. The low mass of dwarf irregular galaxies makes their
ISM particularly vulnerable to disruption by intense episodes of star
formation (starbursts). Starbursts inject turbulence into the
interstellar medium and may provide enough energy to drive an outflow
of gas from the galaxy
\citep{ch2:1999ApJ...513..142M,ch2:2005A&A...436..585D,ch2:2006ApJ...643..186T,ch2:2008ApJ...674..157C}. While
there have been several studies on the effects of starbursts on the
warm \citep{ch2:1998ApJ...506..222M}, hot
\citep{ch2:2005MNRAS.358.1453O}, and cold gas
\citep{ch2:2004ApJ...610..201S} in dwarf irregular galaxies, little
attention has been paid to the role of magnetic fields in starbursting
dwarf irregular galaxies.

The presence of magnetic fields can influence the behavior of the ISM
in these galaxies in several ways.  First, the field is an important
source of pressure for the interstellar medium. Based on observations
of the ISM of the Milky Way \citep[for a comprehensive review
see][]{2001RvMP...73.1031F}, one often assumes that the interstellar
medium of galaxies is in equipartition with approximately equal energy
in the magnetic field, cosmic rays, and turbulence.  Second, given the
shallow gravitational potential well of these galaxies, the magnetic
field may be strong enough to be dynamically important and aid in
expelling gas from the disk of the galaxy. The configuration of the
field, however, may help or hinder the outflow.  Although earlier work
by \citet{ch2:1990ApJ...361L...5T} and \citet{ch2:1993ApJ...409..663M}
came to the conclusion that magnetic fields parallel to the disk could
prevent the breakout of superbubbles from the disk, more recent 3D
simulations by \citet{ch2:2005A&A...436..585D} show that while a disk
magnetic field may delay the break out of superbubbles, the amount of
hot gas vented from superbubbles is not greatly affected by the
addition of a magnetic field. For the case of a magnetic field
perpendicular to the galactic disk,
\citet{ch2:1991A&A...245...79B,ch2:1993A&A...269...54B} present models
of a cosmic-ray driven wind, where the magnetic field transfers
momentum and energy from cosmic rays to thermal gas thus helping to
drive the wind. \citet{ch2:2008ApJ...674..258E,2009arXiv0904.1964E}
later extended these models to explain the distribution of soft
Galactic X-rays and 408~MHz emission towards the Galactic Center. The
cosmic rays, acting through the magnetic field, may increase mass loss
rates or increase the terminal velocity of the wind depending on the
strength and geometry of the magnetic field
\citep{ch2:2008ApJ...674..258E,2009arXiv0904.1964E}.





There are a handful of starburst galaxies with detailed magnetic field
observations including M82
\citep{ch2:1988A&A...190...41K,ch2:1994A&A...282..724R}, NGC 253
\citep{ch2:1992ApJ...399L..59C,ch2:1994A&A...292..409B,2009A&A...494..563H},
NGC 4666 \citep{ch2:1997A&A...320..731D}, NGC 1808
\citep{ch2:1990A&A...240..237D}, and NGC 4569
\citep{ch2:2006A&A...447..465C}. The common features in all these
objects are extended radio continuum emission well beyond the disk of
the galaxy and spurs of polarized radio continuum emission extending
from the disk of the galaxy to the halo of the galaxy with their
derived magnetic field vectors roughly perpendicular to the
disk. However, all these galaxies are quite massive -- NGC 253, NGC
4666, NGC 1808, and NGC 4569 have dynamical masses on the order of
$10^{11} \ \rm{M}_\sun$, while M82 has a dynamical mass on the order
of $10^{10} \rm{M}_\sun$. A low-mass galaxy like a dwarf irregular,
which has a dynamical mass on the order of $10^9\ \rm{M}_\sun$,
provides a more rigorous test of the role of the magnetic field in a
starburst.



One of the most extreme starbursts in the local universe is the dwarf
irregular galaxy NGC 1569. This galaxy, which is only 3.36 Mpc away
\citep{2008ApJ...686L..79G},
underwent an intense burst of star formation (star formation rate
$\sim 1-3 \ M_\odot \ \rm{yr}^{-1} \ \rm{kpc}^{-2}$) in the recent
past ($\sim$ 10--40 Myr ago)
\citep{ch2:1996A&A...313..713V,ch2:1998ApJ...504..725G,ch2:2001AJ....121.1425A,ch2:2005AJ....129.2203A}. Today,
NGC 1569 is in a post-starburst phase, although it is still vigorously
forming stars \citep{ch2:1998ApJ...504..725G}. The subsequent
evolution and demise of the high-mass stars formed in the starburst
has injected an enormous amount of energy into the interstellar medium
of NGC 1569 causing massive disruption of the gas. The \ha structure
of this highly inclined galaxy ($i = 63$;
\citealp{ch2:2002A&A...392..473S}) consists of filaments of emission
extending well into the halo of the galaxy. The most prominent \ha
filament is referred to as the ``western arm'' (note that this is not
a spiral arm, but the limb of an \ha bubble).  Investigations of the
\ha kinematics
\citep{ch2:1994PASJ...46..335T,ch2:1995ApJ...448...98H,ch2:1998ApJ...506..222M,ch2:2007MNRAS.381..894W,ch2:2007MNRAS.381..913W,ch2:2008MNRAS.383..864W}
reveal that the central regions of the galaxy are dominated by
turbulent motions, while the outer regions show clear signs of a
large-scale outflow of ionized gas. The neutral hydrogen velocity
field is completely disrupted in the central regions of NGC 1569, but
rotates as a solid-body in the outer regions
\citep{ch2:2005AJ....130..524M}. X-ray observations
\citep{ch2:1995ApJ...448...98H,ch2:1996ApJ...469..662D,ch2:2002ApJ...574..663M,ch2:2005MNRAS.358.1423O,ch2:2005MNRAS.358.1453O}
have revealed the widespread presence of hot, partially shock-heated,
gas bounded by the \ha filaments throughout the galaxy. This gas has a
high $\alpha$-to-Fe ratio indicating that the hot gas is enriched with
the products of type II supernovae
\citep{ch2:2002ApJ...574..663M}. Although NGC 1569 clearly has gas
being expelled from its central regions and the gas likely has enough
energy to escape
\citep{ch2:1998ApJ...506..222M,ch2:2005MNRAS.358.1453O}, how much
material, if any, will actually escape the galactic potential remains
controversial. A detailed study of the magnetic field strength and
structure of NGC 1569 is a crucial component for understanding the
outflow of gas from this galaxy.

The magnetic field of a galaxy can be traced using its radio continuum
emission, which is a combination of synchrotron radiation generated by
relativistic electrons spiraling along magnetic field lines and
free-free emission generated by thermal electrons. The resolved radio
continuum emission from NGC 1569 has been observed by
\citet{ch2:1976A&A....48..421S}, \citet{ch2:1983ApJS...53..459C},
\citet{ch2:1986A&A...161..155K} (1.2\,cm [24.5\,GHz] data only),
\citet{ch2:1988A&A...198..109I}, and
\citet{ch2:2004MNRAS.349.1335L}. \citet{ch2:1976A&A....48..421S} found
that most of the regions they detected in NGC 1569 are thermal in
nature; however, these observations were only sensitive to a limited
range of spatial scales with little sensitivity to large-scale
emission. \citet{ch2:1983ApJS...53..459C} presented VLA observations
of NGC 1569 showing an extended radio continuum halo at 20\,cm
(1.4~GHz), but did not discuss them extensively. The Effelsberg 100-m
observations of \citet{ch2:1986A&A...161..155K} only resolved NGC 1569
at 1.2\,cm (24.5~GHz). At this wavelength, they detected faint
emission associated with the western arm for the first time at a
frequency higher than 20\,cm (1.4~GHz).
\citet{ch2:1988A&A...198..109I} presented observations of the radio
continuum emission from NGC 1569 at 50\,cm, 20\,cm, and 6\,cm
(0.6~GHz, 1.4~GHz, and 5.0~GHz, respectively) and discuss in detail
the faint extended radio halo around this galaxy first found by
\citet{ch2:1983ApJS...53..459C} and visible in their observations at
both 50\,cm and 20\,cm. They identified a break in the non-thermal
radio continuum spectrum of this galaxy at 3.75\,cm ($8 \pm 1 \
\rm{GHz}$). \citet{ch2:2004MNRAS.349.1335L} modeled the integrated
radio spectrum (including the break) and concluded that the radio
continuum spectrum is likely the result of cosmic-ray escape in a
convective wind, which fits with the picture of NGC 1569 as a galaxy
undergoing a tremendous outflow of material.

In this paper, we present multi-wavelength radio continuum
polarization observations of the post-starburst dwarf irregular galaxy
NGC 1569, which will allow us to determine the strength of the
magnetic field in this galaxy and trace its structure. This data set
represents the most sensitive set of radio continuum observations of
NGC 1569 to date and the first observations of its polarized emission
at 20\,cm, 13\,cm, 6\,cm, and 3\,cm.  We describe our data set in
\S~\ref{ch2:sec:data-reduction} and present the basic properties of
the total intensity and polarized continuum emission of NGC 1569 in
\S~\ref{ch2:sec:prop-radio-cont}. In \S~\ref{ch2:sec:magn-field}, we
investigate the properties of the magnetic field of NGC 1569. We place
our observations in the context of what is known about the
interstellar medium of NGC 1569 in
\S~\ref{ch2:sec:discussion}. Finally, in
\S~\ref{ch2:sec:summary-conclusions}, we present a summary of our work
and our conclusions.

\section{Observations and Data Reduction} \label{ch2:sec:data-reduction}

We observed the radio continuum emission from NGC 1569 at 20, 13, 6,
and 3\,cm. The 20 and 13\,cm observations were taken with the
Westerbork Synthesis Radio Telescope (WSRT)\footnote{ The Westerbork
  Synthesis Radio Telescope is operated by ASTRON (Netherlands
  Foundation for Research in Astronomy) with support from the
  Netherlands Foundation for Scientific Research (NWO).}. The WSRT is
an east-west interferometer consisting of fourteen 25-m antennas. Ten
of the antennas are at fixed positions, while the other four antennas
may be moved along an east-west track.  At 6 and 3\,cm, we used both
the Very Large Array (VLA)\footnote{The VLA is part of the National
  Radio Astronomy Observatory. The National Radio Astronomy
  Observatory is a facility of the National Science Foundation
  operated under cooperative agreement by Associated Universities,
  Inc.}, which is an interferometer with twenty-seven 25-m antennas
arranged in a Y-shape, and the Effelsberg 100-m Radio
Telescope\footnote{ Operated by MPIfR (Max-Planck-Institut f{\"u}r
  Radioastronomie).}, which is a single-dish telescope. The WSRT and
VLA observations provide us with the spatial resolution necessary to
detect small-scale magnetic field structure that would be depolarized
by the large beam of the Effelsberg observations. The Effelsberg
observations are necessary to recover the emission on the largest
spatial scales, which may be resolved out by the VLA at 6 and
3\,cm. At 20 and 13\,cm, the largest angular scales imaged by the WSRT
are approximately $19\arcmin$ and $17\arcmin$, which are both much
greater than the size of the galaxy, so there is very little chance of
large-scale emission from NGC 1569 being resolved out at these
wavelengths. Below we discuss the calibration and imaging of the data
in detail.

\subsection{20\,cm Data} \label{ch2:sec:20cm-data}

The 20\,cm data were taken with the WSRT as part of project R01A/12
(PI: S. M{\"u}hle) on 5--6 and 6--7 November 2000. The project also
included observations of the radio continuum of NGC 1569 at 49\,cm,
but the 49\,cm data were unusable due to extensive radio frequency
interference.  To calibrate the instrumental polarization, each
observing session included an observation of a source with a known
polarization angle and an unpolarized source. These sources were also
used to calibrate the flux density scale. The time on-source for each
day was 12 hours. To maximize the coverage of the {\em u-v} plane, we
used two different array configurations: the 72\,m configuration and
the 36\,m configuration. The DCB backend was configured to produce
eight 10MHz-wide intermediate frequency (IF) channels each with full
polarization products.

\begin{figure}
\centering
\includegraphics{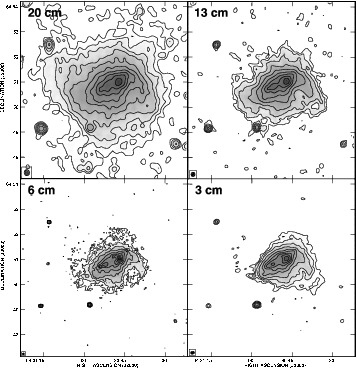}
\caption[Total intensity radio continuum images of NGC 1569]{ 20\,cm,
  13\,cm, 6\,cm, and 3\,cm total intensity, robust=0 weighted
  images. The contours start at $3\sigma$ and increase by factors of 2
  (see Table~\ref{tab:final_image_summary} for the noise levels for
  each image). The beam for each observation appears in the lower left
  corner. See Figure~\ref{ch2:fig:total_intensity_3cm_6cm_only} for a
  more detailed view of the 6\,cm and 3\,cm emission.}
   \label{ch2:fig:total_intensity}
\end{figure}

Table~\ref{tab:wsrt_20cm_obs_summary} summarizes the 20\,cm
observations. The data were calibrated and imaged using the NEWSTAR
package \citep{ch2:1996ASSL..207...53B}, following standard
calibration procedures and using a hybrid cleaning method, which
includes self-calibration of the gains and phases of the Stokes I, U,
and V data. In Stokes Q, only the phases were self-calibrated, since a
self-calibration of the gains would force Stokes Q equal to zero in
the case of linear feeds
(Q=XX-YY). Table~\ref{tab:final_image_summary} gives details for the
final images. The upper left panel of
Figure~\ref{ch2:fig:total_intensity} shows the total intensity map of
the 20\,cm data. The final image includes low-level (20~\uJybeam)
ripples in the total intensity that are artifacts of the imaging
process for this data.



\subsection{13\,cm Data} \label{ch2:sec:13cm-data}

The 13\,cm data were obtained with the WSRT (project ID: R03A/25, PI:
S.\ M{\"u}hle) on 4-5 December 2002 and 18-19 January 2003. The time
on-source for each observing session was 12 hours. Observations of a
polarized source and an unpolarized source were taken before and after
each session to calibrate the instrumental polarization as well as the
bandpasses and the flux density scale.  A summary of the WSRT 13\,cm
observations used in this paper and their calibration is given in
Table~\ref{tab:wsrt_13cm_obs_summary}.  The data were taken in the
36-m and 72-m configurations and DZB backend was configured to produce
eight 20~MHz-wide IFs, each with full polarization products and 64
channels.

Since the 13\,cm receivers on the WSRT have circular feeds, we
calibrated the data using the standard calibration routines in AIPS,
which, in general, assume circularly polarized feeds. Reducing WSRT
data in AIPS has two important differences compared to reducing
typical VLA observations. First, the WSRT is a redundantly spaced
array, so observing phase calibrators periodically during the
observing session is not necessary because the redundant baselines
provide multiple measurements of the same visibility. Self-calibration
is used to solve for the variation of phase as a function of time. The
second complication is that the WSRT dishes are equatorially mounted,
so the parallactic angle does not change as a function of
time. Therefore, a single observation of an {\em unpolarized} source
must be used to calibrate the instrumental polarization. This is in
contrast to the VLA where the polarization response of the antennas is
usually calibrated using a weakly polarized source observed over a
large range in parallactic angle. Practically, it means that the task
LPCAL must be used to calibrate the instrumental polarization instead
of the task PCAL.

We used the AIPS task CALIB to self-calibrate the data using models of
the galaxy generated by imaging and cleaning the data. We did several
iterations of phase-only self-calibration with solution intervals of
between 5 and 10 minutes and restricted the solutions to those
baselines where the signal-to-noise ratio was greater than 3. After
the phases had been improved, we did a final round of amplitude and
phase self-calibration with our best model of the source, which
included as much of the diffuse flux as possible. The solution
interval was between 20 and 30 minutes and only baselines where the
signal-to-noise ratio was greater than 5 were included in the
solution. Each observation period was self-calibrated separately and
the individual periods combined together. Finally, the combined data
was amplitude and phase self-calibrated with a solution interval of 12
hours including only those baselines with signal-to-noise ratio
greater than 5 to ensure a uniform amplitude scale.

We produced two final images: a robust=0 image and an image matched to
the resolution of the high-resolution 20\,cm image ($\sim 13\arcsec$).
The robust=0 image has a weighting between uniform and natural and
provides the best compromise between the lower noise given by natural
weighting and the minimization of the sidelobes by uniform weighting.
Table~\ref{tab:final_image_summary} lists the details of the final
images.  The upper right panel of Figure~\ref{ch2:fig:total_intensity}
shows the robust=0 image.

\subsection{6\,cm \& 3\,cm Data} \label{ch2:sec:6cm--3cm}

The interferometer data at these wavelengths were taken with the VLA
under program IDs AM643 (D array observations) and AM694 (C array
observations). The PI for both programs was S. M{\"u}hle.  A full set
of polarization products (RR, RL, LR, LL) were produced for each of 2
IFs with a bandwidth per IF of 50\,MHz. The total time on-source was
3.97 hours for the 6\,cm (4.8~GHz) D-array observations, 14.7 hours
for the 6\,cm (4.8~GHz) C-array observations, and 10.5 hours for the
3\,cm (8.4~GHz) D-array observations. We lost 2 hours of time on
source during the 3\,cm observations on 23 September 2000 due to high
winds. We calibrated the data in AIPS following the standard
calibration procedures detailed in the AIPS cookbook
\citep{ch2:aipscookbook}. We self-calibrated the data in AIPS using
the same procedure as for the 13\,cm data (see
\S~\ref{ch2:sec:13cm-data}).  Table~\ref{tab:vla_obs_summary}
summarizes our observations and calibration parameters.


Whenever possible, we used 1331+305 (3C286) as our absolute
polarization angle calibrator, since the polarization angles of all
other calibrators are derived from this source. If 1331+305 was not
available, 0521+166 was used as an absolute polarization angle
calibrator. The polarization angle of 0521+166 in the last row of
Table~\ref{tab:vla_obs_summary} was determined using observing
sessions where we observed both 0521+166 and 1331+305. The absolute
polarization angle for 0521+166 on 9 August was determined from
observations on 10 August, on 11 and 17 August from observations on 19
September, on 23 September from observations on that day. Our
observations on 19 September indicate that the absolute polarization
angle of 0521+166 varied by $\pm 2^\circ$ over the entire observing
run.

To correct the 6\,cm and 3\,cm VLA data for the effects of missing
short spacings\footnote{It is not necessary to correct the 20\,cm and
13\,cm data for the effect of missing short spacings since the largest
angular scale imaged by the WSRT at these frequencies is much larger
than the size of the galaxy.}, we obtained 6\,cm (4.85\,GHz) and 3\,cm
(10.45\,GHz) data with the Effelsberg 100-m radio telescope in
December 2000 and January 2001, respectively. The data at both
wavelengths were reduced and calibrated using the NOD2 data reduction
package \citep{ch2:1974A&AS...15..333H}. The 6\,cm (4.85\,GHz)
Effelsberg receiver has two feeds, which allows one to eliminate
atmospheric noise using the dual-beam technique described in
\citet{ch2:1979A&A....76...92E}. The telescope scans in azimuth to
align the two feeds. To remove raster noise, we averaged together
individual maps taken across a range of parallactic
angles. Atmospheric noise at 3\,cm (10.45\,GHz) was minimized by using
the software beam-switching technique on the data from the four feeds
of this receiver. The residual scanning noise in the individual images
was reduced by down-weighting the regions in Fourier space where
raster noise is present \citep{ch2:1988A&A...190..353E}. The
individual maps were then averaged to produce the final
maps. Table~\ref{sec:effelsberg_obs_summary} gives a synopsis of our
Effelsberg observations.

Since the VLA and Effelsberg 3\,cm images were taken at two different
frequencies (8.46 GHz and 10.45 GHz), we scaled the Effelsberg Stokes
I image by the spectral index determined by Lisenfeld et al. (2004):
$\alpha = -0.47$. We did not scale the Effelsberg Stokes Q and U
images because we do not know how NGC 1569's polarized emission
changes between 8.46 GHz and 10.45 GHz. We can, however, constrain the
rotation measure between these two frequencies. For a change of
10$^\circ$ in the angle of the polarization vector, there would have
to be a rotation measure of at least 400 rad~m$^{-2}$. In
\S~\ref{ch2:sec:magn-field-prop}, we determine the rotation measures
for NGC 1569 between 6\,cm and 3\,cm. Most of the measured values are
less than 400 rad~m$^{-2}$ (see Figure~\ref{ch2:fig:rotmeas_6cm_3cm}),
so not scaling the flux in the Effelsberg 10.45~GHz Stokes Q and U
images should not be problematic. However, rotation measures greater
than, or comparable, to 400 rad~m$^{-2}$ may be affected by combining
unscaled Stokes Q and U data taken at two different frequencies.


To combine the interferometric and single-dish data, we used the {\em
  immerge} task in Miriad, which combines the images in the spatial
  frequency domain. In this approach, the different calibration scales
  of the Effelsberg and VLA data are dealt with by scaling the
  low-resolution visibilities so that they match the high-resolution
  visibilities in the region of the $u-v$ plane where the observations
  overlap (0.4 to 1.6~$\rm{k}\lambda$ for the 6\,cm data and 0.725 to
  2.8~$\rm{k}\lambda$ for the 3cm data). The scaling factors for the
  visibilities in the overlap region as determined by {\em immerge}
  were 0.8196 (6\,cm, robust=0 image), 0.856 (6\,cm, 13\arcsec\
  image), 0.9263 (3\,cm, robust=0 image), and 0.9330 (3\,cm,
  13\arcsec\ image). We used the scaling factors determined by the
  Stokes I images for the Stokes Q and U images.

\begin{figure}
\centering
\includegraphics{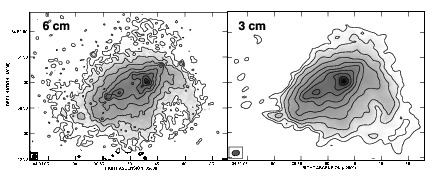}
\caption[Zoomed in version of 6 and 3\,cm total intensity radio
continuum images of NGC 1569]{Zoomed-in version of the 6\,cm and 3\,cm
  total intensity, robust=0 weighted images. The first contour is at
  $3\sigma$ and subsequent contours increase by factors of two (see
  Table~\ref{tab:final_image_summary} for the noise levels for various
  images). The beam appears in the lower left corner of each panel.}
\label{ch2:fig:total_intensity_3cm_6cm_only}
\end{figure}

\begin{figure}
\centering
\includegraphics[scale=0.8]{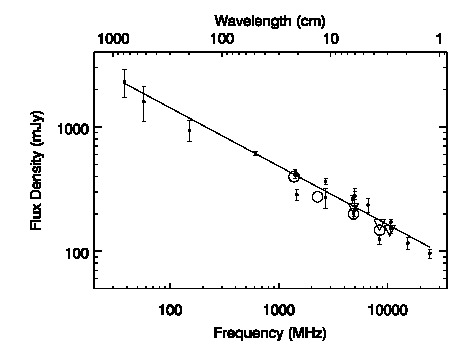}
\caption[Radio continuum spectrum of NGC 1569]{Integrated flux density
  spectrum for NGC 1569. The small dots are from the compilation of
  \citet{ch2:2004MNRAS.349.1335L} and open circles are the total flux
  in our final images.  The error in the total flux of our final
  images is less than the size of the symbol. The open triangles are
  the total flux densities determined from our Effelsberg
  observations. The 8~GHz flux was determined by scaling our 10~GHz
  Effelsberg observations to 8~GHz assuming a spectral index of
  0.47. The error bars show $1\sigma$ errors. The line shows a power
  law with a slope of $\alpha = 0.47$. }
  \label{ch2:fig:total_flux_spectrum}
\end{figure}

Details on the final 6\,cm and 3\,cm images are given in
Table~\ref{tab:final_image_summary}. Figures~\ref{ch2:fig:total_intensity}
and \ref{ch2:fig:total_intensity_3cm_6cm_only} show the robust=0
weighted 6\,cm and 3\,cm images. The total fluxes of the final images
are within 10\% of the total fluxes measured with Effelsberg (see
Figure~\ref{ch2:fig:total_flux_spectrum}).

\section{Properties of the Radio Continuum Emission} \label{ch2:sec:prop-radio-cont}

The total intensity of the radio continuum emission from a source
gives one information about the combined free-free and synchrotron
emission in that source. By separating the synchrotron emission from
the free-free emission, one can estimate the total magnetic field
strength (see \S\S\ref{ch2:sec:estim-therm-flux} and
\ref{ch2:sec:magn-field-prop}). Since free-free emission is
unpolarized and synchrotron emission is polarized, any polarized
emission is inherently synchrotron emission and traces of the uniform
component of the magnetic field (see
\S~\ref{ch2:sec:magn-field-struct}).

\subsection{Total Intensity} \label{ch2:sec:total-intensity}

Previous resolved studies of NGC 1569 at radio continuum wavelengths
include \citet{ch2:1976A&A....48..421S},
\citet{ch2:1983ApJS...53..459C}, \citet{ch2:1986A&A...161..155K}
(1.2\,cm [24.5~GHz] data only), \citet{ch2:1988A&A...198..109I}, and
\citet{ch2:2004MNRAS.349.1335L}. The details of these observations are
shown in Table~\ref{tab:previous_obs}; where necessary, we converted
sensitivities in brightness temperatures to sensitivities in \mJybeam
using the given frequencies and beam parameters. Our observations
reach surface brightnesses two (20\,cm observations) to eighty (13\,cm
observations) times fainter than previous observations. Although our
resolution is not as high as some previous observations, we do have
better {\em u-v} plane coverage, making our observations sensitive to
both compact sources and extended emission.

We have detected radio continuum emission from NGC 1569 at all four
observed wavelengths.  In Figure~\ref{ch2:fig:total_flux_spectrum}, we
plot the total flux detected by our measurements along with radio
continuum flux measurements collected from the literature by
\citet{ch2:1988A&A...198..109I} and
\citet{ch2:2004MNRAS.349.1335L}. Our measured fluxes at 20\,cm
(1.4~GHz) and 13\,cm (2.7~GHz) agree well with the measured fluxes
from the literature, justifying our assumption not to correct the WSRT
observations at these frequencies for the effects of missing
short-spacings. The 6\,cm (4.86~GHz) and 3\,cm (8.46~GHz) fluxes from
our combined VLA and Effelsberg images are -9.5\% (6\,cm) and -8.8\%
(3\,cm) less than the measured Effelsberg fluxes.  (Note that the low
point from the literature at 3\,cm was an interferometer measurement
that did not correct for missing short-spacings.)  All our
observations are at wavelengths greater than or approximately equal to
the break in the radio continuum spectrum at approximately 3.75\,cm
(8~GHz) first noted by \citet{ch2:1988A&A...198..109I} and modeled by
\citet{ch2:2004MNRAS.349.1335L}.

Our total intensity images at all four frequencies are shown in
Figures~\ref{ch2:fig:total_intensity} and
\ref{ch2:fig:total_intensity_3cm_6cm_only}. Note that we show the
highest resolution image at each frequency, so the beam size varies
between the images. At 20\,cm, we detect both the bright central
regions and the extended diffuse component first seen by
\citet{ch2:1983ApJS...53..459C} and first discussed by
\citet{ch2:1988A&A...198..109I}. There are two bright peaks of
emission in the central regions of the galaxy. The brightest peak is
on the northwestern side of the disk and the fainter, slightly
elongated peak is on the southeastern side of the disk. In our 6\,cm,
robust=0 image, which has the highest resolution of all our images, we
resolve the southeastern peak into two sources. Our observations
increase the extent of the diffuse component of the galaxy from $\sim
3\arcmin$ \citep{ch2:1988A&A...198..109I} to $\sim 4\arcmin$. This
corresponds to a physical extent of approximately $4.4\ \mathrm{kpc}$
off the plane of the disk (assuming an inclination of 63$^\circ$;
\citealp{ch2:2002A&A...392..473S}), which is extraordinary considering
the disk is only about 3$\arcmin$ ($2.9 \ \mathrm{kpc}$) long in the
radio continuum. We also detect the western arm of the galaxy at all
observed frequencies, where previous observations only saw it at
20\,cm.\footnote{ \citet{ch2:1986A&A...161..155K} saw a hint of this
  feature in 24.5~GHz single-dish data.} As one goes from 20\,cm to
3\,cm, the morphology of the diffuse component of the radio continuum
emission changes from box-shaped to more oval-shaped. At shorter
wavelengths, the western arm becomes increasingly prominent and the
extent of the radio continuum halo decreases.

Spectral index maps of the observed total intensity radio continuum
emission can provide an important diagnostic of the processes
producing the radio continuum emission. We define the spectral index,
$\alpha$, as $S \propto \nu^\alpha$, where $S$ is the source flux and
$\nu$ is the frequency; thermal emission has a spectral index of
$-0.1$ and synchrotron emission has a spectral index of $\sim
-0.7$. By looking at the observed spectral index, we can get an idea
of whether thermal or synchrotron emission dominates in a particular
region.

Spectral index maps, however, are significantly affected by small
systematic changes in flux. For example, a 5\% change in the flux
level in each total intensity map going into the spectral image can
change the spectral index by as much as 0.18. Because of the artifacts
remaining in the 20\,cm and 13\,cm images and the systematic
uncertainties in creating spectral index maps from data from two
different telescopes (especially ones as different as the VLA and the
WSRT), we created a spectral index map for NGC 1569 based only on the
6\,cm and 3\,cm data. To make sure that both images sampled the same
spatial scale, we used only the interferometric data and restricted
the $u$-$v$ range of the data imaged to 0.74 to 21 k$\lambda$; this
range is the $u$-$v$ range common to the observations at 13\,cm,
6\,cm, and 3\,cm.\footnote{The range common to the 6\,cm and 3\,cm is
  0.74 to 29.3 k$\lambda$, which would increase the resolution
  slightly.} The beam sizes in each image were slightly different
because differences in the sampling of the $u$-$v$ plane in each
image. We smoothed both images to the resolution of the image with the
largest beam (9.34\arcsec\ by 7.54\arcsec).

\begin{figure}
\centering
\includegraphics[angle=-90,scale=0.5]{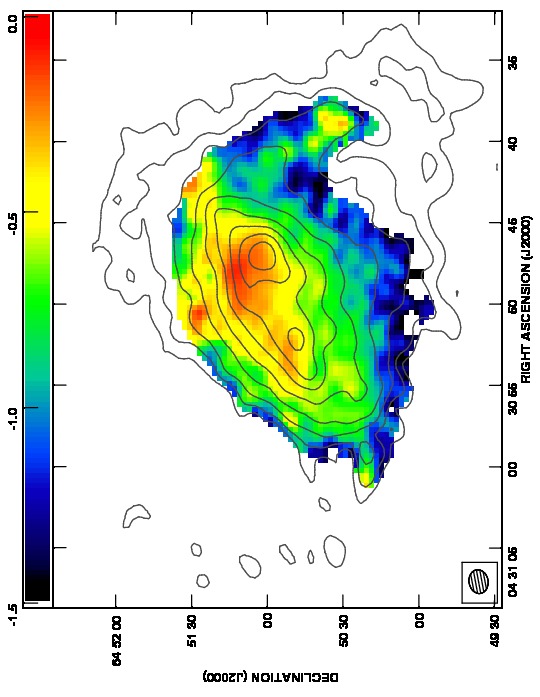}
\caption[Spectral indices between adjacent wavelengths for NGC 1569]{
  Spectral index map between 6\,cm and 3\,cm data. For this map, both
  data sets were imaged with the same $u$-$v$ range (0.74 to 20
  k$\lambda$) and smoothed to the same beam size (9.34\arcsec\ by
  7.54\arcsec, shown in the lower left corner). Regions where the
  formal error on the spectral indices exceeded 0.2 were blanked. The
  contours show the 3\,cm, robust=0 image and start at the $3\sigma$
  noise level and increase by a factor of two for each subsequent
  contour. }
  \label{ch2:fig:spectral_index_plot}
\end{figure}

Figure~\ref{ch2:fig:spectral_index_plot} shows the spectral indices
between the 6\,cm and 3\,cm data. Only spectral indices where the
formal error on $\alpha$ was less than 0.2 were used. The formal error
on $\alpha$ is
\begin{equation}
\sigma_\alpha^2 = \left[\frac{1}{\log(\nu_1/\nu_2)}\right]^2
\left[\left(\frac{\sigma_1}{S_1}\right)^2 
+ \left(\frac{\sigma_2}{S_2}\right)^2 \right]
\end{equation}
where $S$, $\nu$, and $\sigma$ are the flux in a pixel, the frequency,
and the noise for a particular image and the subscripts denote the
first and second images. The values in our final spectral index map
agree with those in a spectral index map created using data that was
not self-calibrated, showing that our self-calibration of the 6\,cm
and 3\,cm data has not introduced major uncertainties into our spectral
index maps.

The spectral indices in the galactic disk are close to thermal with
flattest spectral indices offset by 14\arcsec\ (a little more than a
beam) to the north from the main ridge of 3\,cm continuum emission.
Since the southern half of the galaxy is pointed toward the observer
and the northern half pointed away \citep{ch2:2002ApJ...574..663M},
this spectral index asymmetry about the major axis of the disk can be
explained as free-free absorption of synchrotron emission from behind
the disk by the disk itself. Free-free absorption flattens spectral
indices \citep{ch2:1992ARA&A..30..575C}. In general, as one moves
along the minor axis of the galaxy, away from the disk, the spectral
indices become steeper, eventually reaching values of approximately
--1 at the southern edge of the galaxy. There are, however, two
filamentary regions of flat spectral indices extending to the outer
edge of the spectral index map. These flat spectrum filaments are
aligned with the outer edges of an \ha bubble (Bubble B in the
terminology of \citealt{ch2:2008MNRAS.383..864W}, see also
Figure~\ref{bubble_finding_chart}). The spectral index trends seen
here agree with the general trends seen in the higher resolution data
of \citet{ch2:2004MNRAS.349.1335L} supporting their theory that the
break in the total non-thermal emission spectrum of NGC 1569 is due to
the convective transport of cosmic rays into the halo.

\subsection{Polarized Emission} \label{ch2:sec:polarized-emission}

\begin{figure}
\centering
\includegraphics{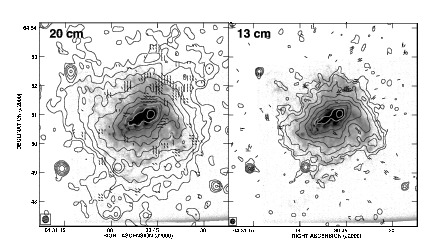}
\caption[Polarized intensity at 20\,cm and 13\,cm of NGC 1569]{
  E-vectors showing the direction and distribution of the polarized
  intensity at 20\,cm and 13\,cm with contours showing the total
  intensity radio continuum emission at each wavelength (robust=0
  weighting) and a grayscale image showing the distribution of the \ha
  emission.  A polarization vector with a length of $5 \arcsec$ has a
  polarized intensity of 51.1 $\mu \rm{Jy} \ \rm{beam}^{-1}$. The
  contour levels are for the robust=0 image in each case. They start
  at $3\sigma$ and increase by a factor of 2 for subsequent contour
  (see Table~\ref{tab:final_image_summary} for the noise levels for
  each image).  The central portion of NGC 1569 has been saturated in
  the grayscale image to better show the faint, extended emission. The
  beams are indicated the lower left hand corner of each panel. The
  line in the greyscale image roughly running from RA
  $04^{\rm{h}}30^{\rm{m}} 45^{\rm{s}}$ to $04^{\rm{h}} 30^{\rm{m}}
  22^{\rm{s}}$ at a declination of 64\degr 47\arcmin\ is due to the
  edge of the CCD.}
\label{ch2:fig:polarized_intensity_20cm_13cm}
\end{figure}

\begin{figure}
\centering
\includegraphics{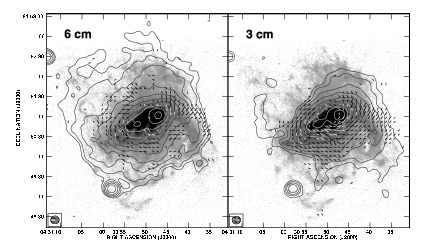}
\caption[Polarized intensity at 6\,cm and 3\,cm of NGC 1569]{
  E-vectors showing the direction and distribution of polarized
  intensity at 3\,cm and 6\,cm with contours showing the total
  intensity radio continuum emission at each wavelength and a
  greyscale image showing the distribution of the \ha emission.  We
  show the 6\,cm and 3\,cm data with $13\arcsec$ resolution because
  these were the data with the most polarized intensity. The contours
  start at $3\sigma$ and increase by a factor of 2 for each subsequent
  contour.  A polarization vector with a length of $1 \arcsec$ has as
  polarized intensity of 12.8~$\mu \rm{Jy} \ \rm{beam}^{-1}$.  The
  beams are indicated in the lower left corner of each panel.}
\label{ch2:fig:polarized_intensity_3cm_6cm}
\end{figure}

The observations presented in
Figures~\ref{ch2:fig:polarized_intensity_20cm_13cm} and
\ref{ch2:fig:polarized_intensity_3cm_6cm} are the first published
observations of the polarized emission of NGC 1569. We detect
polarization at 20\,cm (36 to 80 \uJybeam), 13\,cm (49 \uJybeam to 83
\uJybeam), 6\,cm (30 to 108 \uJybeam), and 3\,cm (31 to 137
\uJybeam). At 3\,cm and 6\,cm, there is polarized emission over a
large portion of the galaxy, while at 20\,cm the polarized emission is
much spottier and at 13\,cm there are only a few small patchies of
polarized emission. Note that the strongest polarization detections at
6 and 3\,cm were with the images with 13\arcsec\ resolution, i.e.,
similar resolutions to the 20\,cm
data. Figures~\ref{ch2:fig:polarized_intensity_20cm_13cm}
and~\ref{ch2:fig:polarized_intensity_3cm_6cm} show the distribution
and direction (E-vectors) of the polarized intensity at the observed
wavelengths.\footnote{We plot E-vectors in
  Figures~\ref{ch2:fig:polarized_intensity_20cm_13cm}
  and~\ref{ch2:fig:polarized_intensity_3cm_6cm} because they are the
  quantity actually measured by the observations. We rotate the
  E-vectors by $90\arcdeg$ to estimate the direction of the magnetic
  field in Figure~\ref{ch2:fig:rotmeas_6cm_3cm}.} To correct for the
bias in the estimated polarized intensity, we use the POLCO task in
AIPS with a signal-to-noise cut of three. This signal-to-noise cut
ensures that all polarization vectors are significant at the 99\%
confidence level
\citep{ch2:simmons1985,ch2:2006PASP..118.1340V}. Decreasing the
signal-to-noise cut to two did not add appreciably more
vectors. Despite NGC 1569's position near the Galactic plane, $(l,b)=
(143\fdg7, 11\fdg2$), there is fortunately very little foreground
Faraday rotation along this line of sight. From the Galactic Faraday
rotation map of \citet{ch2:2004mim..proc...13J}, we estimate that the
foreground Faraday rotation is approximately $+10\ \rm{rad} \
\rm{m}^{-2}$. Examining the polarized sources in our 20\,cm and 13\,cm
data (unfortunately, the only polarized source in our 6\,cm and 3\,cm
data is NGC 1569 itself), we find an average rotation measure between
the 20\,cm and 13\,cm data of --8.7~$\rm{rad} \ \rm{m}^{-2}$ with a
standard deviation of 36.7 $\rm{rad} \ \rm{m}^{-2}$, which agrees with
the value from the \citet{ch2:2004mim..proc...13J} map within the
errors.



The polarized emission at 20\,cm is shown in the left hand panel of
Figure~\ref{ch2:fig:polarized_intensity_20cm_13cm}. The percent
polarization in this image ranges from 2\% at the center to 70\% at
the edges. The central region of NGC 1569 is bright enough to have
significant polarized emission resulting from residual instrumental
leakages rather than emission intrinsic to the galaxy. Based on
measurements of unpolarized sources in the field, we estimate this
leakage at approximately 1\%. Therefore, we require the polarized
emission at 20\,cm to have a fractional polarization of greater than
1\% as well as a signal-to-noise ratio of three to be considered
significant.  The emission in the final image could be depolarized if
the polarization vectors rotate significantly over the observing
frequencies that are combined to produce the final image; this effect
is known as bandwidth depolarization. To determine if the 20\,cm data
is affected, we imaged the first and last intermediate frequencies
(IFs) separately (see Table~\ref{tab:wsrt_20cm_obs_summary} for a
summary of our observing setup). These two IFs are each 10~MHz wide
and 70~MHz apart. Comparing the polarization vectors for the two IFs,
we do not see any rotation of the polarization vectors greater than a
few degrees. Therefore, creating the final image by combining the data
from all the IFs has not resulted in significant depolarization.

\begin{figure}
\centering
\includegraphics[scale=0.5]{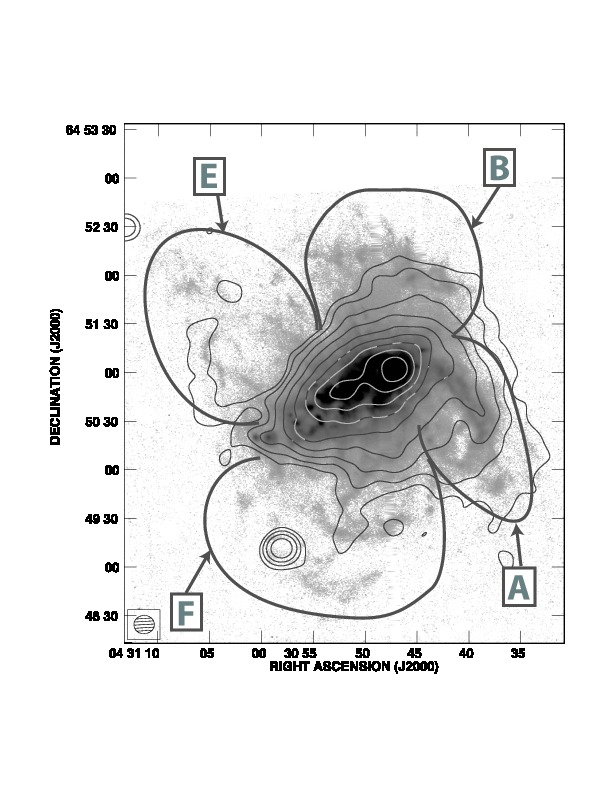}
\caption{An \ha image image from \citet{ch2:hunter2004} with outlines
  indicating the \ha bubbles from \citet{ch2:1998ApJ...506..222M} and
  \citet{ch2:2008MNRAS.383..864W} and contours showing the 3\,cm radio
  continuum emission with $13\arcsec$ resolution. The contours start
  at $3\sigma$ and subsequent contours increase by factors of two. See
  Table~\ref{tab:final_image_summary} for the noise levels in each
  image and Figure 2 in \citet{ch2:2008MNRAS.383..864W} for the
  correspondence between this nomenclature and the nomenclature of
  \citet{ch2:1993AJ....106.1797H}. }
  \label{bubble_finding_chart}
\end{figure}

The major polarization features at 20\,cm are associated with \ha
features in the galaxy.  Figure~\ref{bubble_finding_chart} shows the
positions of the various \ha bubbles that have been identified in the
literature. The \ha emission is included in both
Figure~\ref{ch2:fig:polarized_intensity_20cm_13cm} and
\ref{bubble_finding_chart} as a point of comparison between the
figures. There are polarization features on the northwest and
southwest edges of the disk which correspond to the bases of bubbles B
and A. Moving northwest away from the disk there are regions of
significant polarized emission that outline the edges of bubble B as
well as some that may be associated with the region of \ha filaments
just to the east of this feature.  We also see significant polarized
emission outlining the bubble A.

The 13\,cm emission is shown in the right panel in
Figure~\ref{ch2:fig:polarized_intensity_20cm_13cm}. The polarization
ranges from 1.4\% for vectors near the center of the image to 42.1\%
for the clump of significant vectors in the northwestern portion of
the galaxy.  We used one of the calibration sources (3C147) to
estimate the residual instrumental leakages at 13\,cm as 0.5\% and
eliminated all vectors with smaller percent polarizations. This cut
affected only one vector located on the brightest peak of 13\,cm
emission.  There is very little significant polarization in the 13\,cm
data besides the polarization vectors in the northwestern portion of
NGC 1569 that correspond to locations of significant polarization
vectors at 20\,cm. The lack of significant polarized emission is not
surprising since the 13\,cm has the highest noise level of all our
data. When the 20\,cm polarized intensity is calculated assuming a
noise cutoff similar to that of the 13\,cm data we see little
polarization, which confirms that the lack of polarization at 13\,cm
is most likely the result of the relatively poor sensitivity of this
data. To make sure this is not the result of bandwidth depolarization,
we have imaged first and last 20~MHz IFs by themselves and compared
polarized emission in the two images. Neither one of the single IF
images shows any significant polarization, so significant polarization
in one IFs is not canceling out significant polarization in another
IF.






The most polarized emission is seen in the 6\,cm and 3\,cm data
(Figure~\ref{ch2:fig:polarized_intensity_3cm_6cm}). The percent
polarization seen in the 6\,cm and 3\,cm data is well above the
estimated leakage of the VLA ($<$1\% near the center of the field;
\citealp{ch2:obsstatus}). Therefore, the only cut we have applied to
this data is to require the signal-to-noise ratio for the polarized
emission to be greater than three. At both 6\,cm and 3\,cm, there is
no visible rotation of the polarization vectors between the two IFs,
so bandwidth depolarization does not seem to be an issue. There are
some significant polarization vectors outside the 3$\sigma$ contour
line for the 3\,cm total intensity. For the 3\,cm data with the
$13\arcsec$ resolution, however, the error in the Stokes I image is
larger than estimates of the theoretical noise in Stokes I based on
the noise in the Stokes Q and U images. Therefore, we are relatively
more sensitive to polarized emission than to total intensity emission.



In general, the polarized emission at 3\,cm is strong and present over
a large fraction of the galaxy. The polarized emission at this
wavelength goes from 1-2\% at the center to up to 40\% and greater on
the western edge and 10-30\% in the halo of the galaxy. There is
little and/or weak polarized emission coincident with the western \ha
arm and another break in the emission at approximately $04^{\rm{h}}
30^{\rm{m}} 45^{\rm{s}}$, $64^\circ 50\arcmin 20\arcsec$. The central
regions of the 3\,cm data appear to be largely unpolarized except for
some polarized emission associated with the main radio continuum
emission peak. This depolarization could be due either to Faraday
depolarization or to an inherently turbulent field. We discuss these
possibilities further in \S~\ref{ch2:sec:discussion} in the context of
NGC 1569's \ha kinematics. The majority of the polarized emission is
in the southern half of the galaxy, which is unsurprising given that
X-ray observations have found that the southern portion of the galaxy
is inclined towards us and the northern part is inclined away (see
Figure 13 in \citealt{ch2:2002ApJ...574..663M}). The polarized
emission from the northern half of the galaxy has a longer path length
through ionized gas and thus is subject to more internal
depolarization. There is particularly strong polarization on the
western edge of the \ha arm. This feature is also seen in higher
resolution images not included here. The region on the interior of the
\ha arm (Bubble A in Figure~\ref{bubble_finding_chart}) is polarized
as well as the base of bubble F to the southeast of the disk. In the
northern half of the galaxy, there is polarized emission on the
northeastern edge corresponding to Bubble E and to the north of the
main radio continuum peak, which corresponds to Bubble B.

The polarized emission at 6\,cm is patchier than at 3\,cm and ranges
from 1-2\% in the inner regions of the galaxy to 15-20\% in the outer
regions of the galaxy to greater than 40\% in the polarized
arm. Interestingly, the E-vectors in common between the 3\,cm and
6\,cm data show little sign of Faraday rotation except for the
polarization in the northern half of the galaxy where there is
significant rotation (see \S~\ref{ch2:sec:magn-field-struct}). There
is significantly more polarized emission north of the plane of the
galaxy at 6\,cm than at 3\,cm. Since emission at longer wavelengths
tends to trace emission at smaller Faraday depths, this emission is
probably closer to the observer along the line of sight than the
emission at 3\,cm. At 6\,cm there, is still polarized emission
surrounding the \ha arm as well as polarized emission at the base of
Bubble F in a region of extensive \ha filaments and in the Bubble E
region. These regions of emission, however, are less extended than
those seen in the 3\,cm data. The central region of NGC 1569 at 6\,cm
has no significant polarization.

In short, our observations reveal that more polarized emission is seen
in NGC 1569 at shorter wavelengths than at longer wavelengths and that
the polarized emission is associated with \ha bubbles and filaments in
the galaxy. The first result is common to radio continuum polarization
observations of other galaxies
\citep{1996ARA&A..34..155B}. Depolarization due to Faraday dispersion
and differential Faraday rotation increases with increasing
wavelength, so radio continuum polarization observations of galaxies
at longer wavelengths show less polarization. We defer discussion of
the second result to \S~\ref{ch2:sec:magn-field-struct}.

\section{Magnetic Field Structure and  Strength} \label{ch2:sec:magn-field}

\subsection{Estimating the Synchrotron Flux} \label{ch2:sec:estim-therm-flux}

To determine the magnetic field strength in NGC 1569 and thus assess
the relative importance of the field to the ISM, we must estimate the
portion of the total radio continuum flux due to synchrotron emission
and the portion due to free-free emission. To do this, we estimated
the thermal component of the radio continuum emission using the \ha
emission from the galaxy. We used a flux calibrated \ha image kindly
provided by Deidre Hunter \citep{ch2:hunter2004} to determine the \ha
flux in each pixel. We corrected the resulting \ha flux for Galactic
extinction using a value of $A_{H\alpha} = 1.26$
\citep{ch2:relano2006}.  Since the internal extinction of NGC 1569 is
small ($A_{H\alpha} < 0.6 \ \rm{mag}$) and largely confined to the
central regions of the galaxy \citep{ch2:relano2006}, we do not
attempt to correct for internal extinction. Not correcting for
internal extinction will cause us to underestimate the \ha flux by, at
most, a factor of 1.74 in the regions with the greatest extinction,
which are the region surrounding the central peak and the region
connecting the two brightest radio continuum peaks (see Figure 9 of
\citealt{ch2:relano2006}). Note that we do not need to take into
account non-photoionization processes since the calculation of the
conversion between \ha emission and free-free emission does not depend
on the ionization source.


We calculated the expected thermal emission at our observing
frequencies from the given \ha flux using the expressions given in
Appendix A of \citet{ch2:2004ApJ...606..853H},\footnote{ See
  also \citet{ch2:1986A&A...155..297C} and
  \citet{ch2:1992ARA&A..30..575C}.} which explicitly include the
contribution of ionized helium:
\begin{equation} \label{ch2:eq:fth}
\left( \frac{f_\nu} {\rm{mJy}} \right) = 1.16 \left( 1 +
\frac{n(He^+)}{n(H^+)} \right) \left( \frac{T}{10^4
  \ \rm{K}}\right)^{0.617} \left( \frac{\nu} {\rm{GHz}} \right)^{-0.1}
\left( \frac{f_{H\alpha}} {10^{-12} \ \rm{erg \ cm^{-2} \ s^{-1}} }
\right).
\end{equation}
In Equation~(\ref{ch2:eq:fth}), $f_\nu$ is the radio continuum flux,
$n(He^+)/n(H^+)$ is the ratio of ionized helium to ionized hydrogen,
$T$ is the temperature of the region, $\nu$ is the frequency of the
radio continuum observations, and $f_{H\alpha}$ is the
Galactic-extinction-corrected \ha flux. We use a $n(He^+)/n(H^+)$
ratio of 0.087 \citep{ch2:1997ApJ...483..698M}. The expressions from
\citet{ch2:2004ApJ...606..853H} are valid for temperatures between
10,000K and 20,000K and densities between 100~$\rm{cm}^{-3}$ and
1000~$\rm{cm}^{-3}$. We assume a temperature of 10,000K, which is
typical of HII regions. This temperature is 10\% lower than the lowest
temperature that \citet{ch2:1997ApJ...482..765D} was able to measure
for selected regions of NGC 1569. If we increase the temperature to
11,100~K, this increases the estimated thermal emission by only 7\%,
since the temperature dependence in this equation goes as
$T^{0.617}$. \citet{ch2:2008MNRAS.383..864W} shows that the densities
of ionized gas in NGC 1569 range from $100\ \rm{cm}^{-3}$ (the lowest
density probed by the their diagnostic) to $1000\ \rm{cm}^{-3}$. We
smoothed the result of Equation (\ref{ch2:eq:fth}) to the resolution
of the observations using the AIPS task {\em CONVL}. The synchrotron
flux is
\begin{equation}
F_{nth} = F_{tot}  \left( 1 - \frac{F_{th}}{F_{tot}}\right)
\end{equation}
where $F_{nth}$ is the flux of the synchrotron (non-thermal) emission,
$F_{tot}$ is the total flux observed, and $F_{th}$ is the thermal flux
calculated using Equation~(\ref{ch2:eq:fth}).

\begin{figure}
\centering
\includegraphics{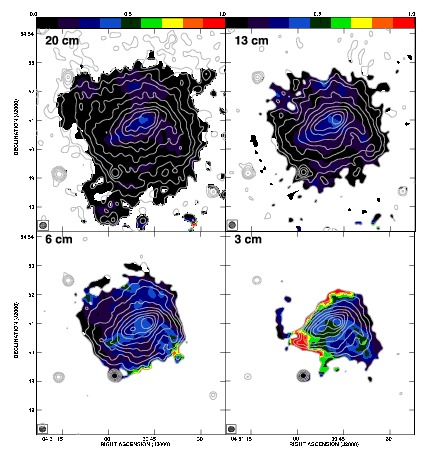}
\caption[Estimated thermal fraction of NGC 1569]{ Fraction of the
  total emission at 20\,cm, 13\,cm, 6\,cm, and 3\,cm that is
  thermal. The thermal emission was estimated using an \ha image of
  the galaxy (see \S~\ref{ch2:sec:estim-therm-flux}). Overlaid are the
  contours of total intensity at each wavelength with $13\arcsec$
  resolution. The contours start at $3\sigma$ and increase by a factor
  of 2 for each subsequent contour (see
  Table~\ref{tab:final_image_summary} for the noise levels for each of
  the images). The beam is shown in the lower left hand corner of each
  panel. The thermal fraction image at 20\,cm does not extend to the
  boundaries of the 20cm emission because the 20\,cm radio continuum
  emission extends beyond the \ha image boundaries, particularly on
  the east and north sides of the galaxy.}
  \label{ch2:fig:thermal_fraction_plot}
\end{figure}

We plot the estimated thermal fraction ($F_{th}/F_{tot}$) in
Figure~\ref{ch2:fig:thermal_fraction_plot}. Note that we set thermal
fractions greater than one to one and thermal fractions less than zero
to zero. This does not significantly change the resulting images and
only affects regions where the emission is faint, i.e., the edges of
the map. As one would expect, the total emission at both 20\,cm and
13\,cm is dominated by synchrotron emission, i.e., has a low thermal
fraction, with the maximum thermal fraction (40\%) southeast of the
location of the radio continuum peak.  The halo of the galaxy has
significantly more synchrotron (less thermal) emission than the
disk. As we go to shorter wavelengths, the amount of thermal emission
increases to 50\% in the region southeast of the radio continuum peak.
Note that the inner portion of the western arm has a significant
amount of thermal emission, while the outer portion is largely
synchrotron emission. The outline of Bubble A appears very clearly in
the 3\,cm map of the thermal emission, which provides additional
confirmation that we are able to accurately determine the location of
the thermal flux with the procedure described earlier in this
section. The entirely thermal regions on the northern and eastern
edges of the 3\,cm map are due to the 3\,cm diffuse emission being
slightly underestimated there. Although we have been very careful
about reconstructing most of the diffuse flux, it is impossible to
reconstruct it all, which results in the thermal fraction being
overestimated in these regions.


\subsection{Magnetic Field Strength} \label{ch2:sec:magn-field-prop}

The intensity of the non-thermal (synchrotron) emission can be used to
estimate the magnetic field strength in a galaxy. The intensity of
synchrotron emission at a particular frequency ($I_\nu$) goes as
\begin{equation}
  \label{ch2:eq:synch_flux}
  I_\nu \propto l N_0 B_\perp^{1 - \alpha } \nu^{\alpha}
\end{equation}
where $l$ is the line of sight through the source, $N_0$ is the
scaling factor for the electron energy spectrum ($N(E) \, \rm{d}E =
N_0 E^{-\gamma} \, \rm{d}E, \gamma = 1- 2\alpha$), $B_\perp$ is the
magnetic field strength in the plane of the sky, and $\nu$ is the
observing frequency \citep{ch2:1965ARA&A...3..297G}.\footnote{We use a
  sign convention for $\alpha$ opposite to the one typically used.}
For a particular value of $I_\nu$ only the quantity $N_0 B_\perp^{1 -
  \alpha}$ can be estimated, so a particular synchrotron intensity may
correspond to a large electron density and a small magnetic field or
vice versa. To disentangle the magnetic field strength, one
traditionally employs the minimum energy assumption where the total
energy (the energy of the magnetic field plus the energy of the cosmic
ray electrons plus the energy of the cosmic-ray protons) is
minimized. For the typical range of $\gamma$, the minimum energy is
similar to the energy derived assuming equipartition between cosmic
rays and the magnetic field \citep{ch2:2005AN....326..414B}. The
minimum energy/equipartition calculation requires a further assumption
that the energy spectrum of the cosmic-ray protons is the energy
spectrum of the cosmic-ray electrons scaled by a constant. However,
this is not true in the Milky Way, where the electron spectrum is
steeper than the proton spectrum (see Figure~1 in
\citealp{2006astro.ph..7109H}).  \citet{ch2:2005AN....326..414B}
present a revised equipartition estimate of the magnetic field
strength that uses the number density ratio of electrons and protons
rather than the more typical energy density ratio to estimate the
field strength. They use the equation below (Equation 3 in
\citealp{ch2:2005AN....326..414B}) to estimate the total equipartition
magnetic field strength:
\begin{equation}
  \label{ch2:eq:beq}
  B_{t} = \left[ \frac{ 4 \pi \ (1 - 2 \alpha) \ ( K_0 + 1) \ I_\nu \
      E_p^{1 + 2\alpha} \ (\nu/2c_1)^{-\alpha} }
{ (-2\alpha-1) \ c_2 \ l \ c_4(i)} \right]^{1/(3 - \alpha)}
\end{equation}
where $B_{t}$ is the total equipartition magnetic field in G, $K_0$ is
the number density ratio of protons to electrons, $E_p$ is the rest
energy of the proton, $c_1$ is a constant equal to $6.26428\times
10^{18} \ \rm{erg}^{-2} \ \rm{s}^{-1} \ \rm{G}^{-1}$, $c_2$ is a
constant tabulated on page 232 of \citet{ch2:1970ranp.book.....P} as
$c_5$, $l$ is the path length through the galaxy, and $c_4(i)$ is a
function correcting for the inclination of the region with respect to
the sky.\footnote{ We use the opposite sign convention for $\alpha$ as
\citet{ch2:2005AN....326..414B}.}  \citet{ch2:2005AN....326..414B}
caution that this formula may not be strictly applicable to starburst
galaxies, since synchrotron losses are important in high-density
regions, and that detailed modeling is needed to determine the true
magnetic field of a starburst. However, based on the work of
\citet{2006ApJ...645..186T}, we believe that our magnetic field
strength estimates for NGC 1569 will not be significantly
underestimated.  \citet{2006ApJ...645..186T} estimate, for a large
number of galaxies from normal spirals to ULIRGs, what they call an
``equipartition'' magnetic field using the assumption that the
magnetic energy density is comparable to the gravitational pressure of
the ISM and compare it to a minimum energy magnetic field estimated
using a method similar to Equation~(\ref{ch2:eq:beq}). The Thompson et
al.\ ``equipartition'' field is the largest magnetic field a galaxy
can have and still be gravitationally stable object. In Thompson et
al.\ Figure~1, NGC 1569 lies on the locus where both the
``equipartition'' magnetic field and the minimum energy magnetic field
are similar, far from the position of galaxies like NGC 253, M82, and
Arp 220, whose minimum energy magnetic field estimates are well below
the Thompson et al.\ ``equipartition'' field.



We use the method of ``magnetic maps'' described in
\citet{ch2:2008A&A...482..755C} to make a map of the total ($B_{t}$),
uniform ($B_{u}$), and random ($B_{r}$) magnetic field strengths as
functions of position in the galaxy along with the revised
equipartition equations given in \citet{ch2:2005AN....326..414B} to
calculate the magnetic field strength.  Equation~(\ref{ch2:eq:beq})
assumes that the magnetic field is either all uniform or all
random. Since the magnetic field of NGC 1569 is clearly a mixture of
the two in many regions, we would like to derive a version of
Equation~(\ref{ch2:eq:beq}) that accounts for both the uniform and
random field components. Equation~(\ref{ch2:eq:beq}) is derived from
the more general Equation A.16 in \citet{ch2:2005AN....326..414B},
which gives the cosmic ray energy density. By equating the magnetic
field energy density ($\epsilon_{B} = B_t^2/8\pi$) to the cosmic ray
energy density ($\epsilon_{CR}$), one obtains
\begin{eqnarray} 
\epsilon_B & = & \epsilon_{CR} \\ 
\frac{B_t^2}{8 \pi}
& = & \frac{ (1 - 2 \alpha) \ (K_0 + 1) \ I_\nu \  E_p^{1+2\alpha}\  (\nu/(2
  c_1))^{-\alpha}}{2 \  (-1 - 2 \alpha) \  c_2 \  l \  B_{\perp}^{(1-\alpha)}}  \\
B_t^2 & = & \frac{ 4 \pi \ (1 - 2 \alpha) \ (K_0 + 1) \ I_\nu \ E_p^{1+2\alpha}\ (\nu/(2
  c_1))^{-\alpha}}{ (-1 - 2 \alpha) \ c_2 \  l \ B_{\perp}^{(1-\alpha)}} \label{eq:b_eq_gen}
\end{eqnarray} 
Note that $\alpha$ here equals $\gamma = 1 - 2 \alpha$ in
\citet{ch2:2005AN....326..414B} and that $c_2$ is a function of
$\alpha$. Since Equation~\ref{eq:b_eq_gen} includes $B_t$, $B_\perp$,
and $l$, it not only needs to take into account the geometry of NGC
1569 and its associated magnetic field, but also the contribution to
the total field from both the uniform and random components. We model
the geometry of NGC 1569 as an infinite cylinder rotated on the plane
of the sky and inclined with respect to the line of sight. Appendix
\ref{ch2:sec:deriving-line-sight} gives an equation for the line of
sight distance through the galaxy.  The geometrical correction to
obtain the total field from the observed field is more complex.  For
an isotropic random field, the correction factor ($c_4$) is
\begin{equation} \label{eq:c4_ran}
c_4 = \left( \frac{B_{r,\perp}}{B_r} \right)^{(1-\alpha)} = \left(\frac{2}{3}\right)^{(1-\alpha)/2}
\end{equation} 
For the regions of NGC 1569 where we do not see any polarization, we
assume that the field is completely random. In this case, the field
strength is given by Equation~(\ref{ch2:eq:beq}) with the geometrical
correction factor given by Equation~(\ref{eq:c4_ran}). For a
completely uniform field,
\begin{equation} \label{eq:c4_uni}
c_4 = \left( \frac{B_{u,\perp}}{B_u} \right)^{(1-\alpha)} = \left[
  (\sin{(\rm{PA}_B)})^2 + \left(\frac{ \cos{(\rm{PA}_B)} } { \sin{i}}\right)^2
  \right] ^ {- ( 1 - \alpha ) / 2}
\end{equation}
where $\rm{PA}_B$ is the position angle of the magnetic field on the
sky. For the rest of this paper, $c_4$ will refer to the version for a
completely uniform field. For a derivation of this quantity, see
Appendix~\ref{ch2:sec:deriving-c_4}.

We cannot assume that regions of polarized emission are completely
uniform; there will still be a contribution of the random component in
these regions. To deal with this, we first need to estimate the ratio,
$q$, of the random magnetic field ($B_{r}$) to the uniform magnetic
field in the plane of the sky ($B_{u,\perp}$). To estimate the ratio
$q$, we use the observed percent polarization ($p$) and the relation
from \citet{ch2:1998MNRAS.299..189S}
\begin{equation} \label{ch2:eq:q}
p = p_0 \frac{1 + (7/3) q^2} {1 + 3 q^2 + (10/9) q^4},
\end{equation}
where $p_0$ is the intrinsic polarization of synchrotron emission
($p_0 =(3-3\alpha)/(5-3\alpha)$).\footnote{
  \citet{ch2:2005AN....326..414B}'s version of this equation has a
  sign error, cf.\ \citet{ch2:1965ARA&A...3..297G}.} Equation
(\ref{ch2:eq:q}) is derived assuming that the cosmic rays are coupled
to the magnetic field
\citep{ch2:1998MNRAS.299..189S,2003A&A...411...99B}. The more commonly
used version of this formula \citep{1966MNRAS.133...67B}
\begin{equation} \label{ch2:eq:q_std}
p = p_0 \frac{ 1}{1 + \frac{2}{3} q^2}
\end{equation}
 does not include this assumption. Both the Equations~(\ref{ch2:eq:q})
and (\ref{ch2:eq:q_std}) assume isotropic fluctuations. Using $q$, one
can now recast the total magnetic field $B_t$ in terms of the $B_{r}$
and $B_{u}$. 
Using the relations $B_{t}^2 =
B_{r}^2 + B_{u}^2$, $B_{\perp}^2 = B_{r,\perp}^2 +
B_{u,\perp}^2$ and the geometrical parameters given in
Equations~\ref{eq:c4_ran} and~\ref{eq:c4_uni},
\begin{eqnarray} \label{ch2:eq:bran}
B_{r,\perp}^2 & = & \frac{2}{3} B_{r}^2 \\
B_{u,\perp} & = & B_{u} c_4^{1/(1-\alpha)}   \\
B_{r} & = & q B_{u,\perp} 
\end{eqnarray}
so
\begin{eqnarray} \label{ch2:eq:btot_comp}
B_{t}^2 & = & ( q^2 c_4^{2/(1-\alpha)} + 1) B_{u}^2 \\ 
B_\perp^2 & = & (\frac{2}{3} q^2 + 1) c_4^{2/(1-\alpha)} B_u^2 \\
B_{\perp}^{(1-\alpha)} & = & (\frac{2}{3} q^2 + 1)^{(1-\alpha)/2} c_4 B_u^{1-\alpha}
\end{eqnarray}
Substituting the above expressions for $B_{t}^2$ and
$B_{\perp}^{(1-\alpha)}$ into Equation~(\ref{eq:b_eq_gen}), we obtain
an expression giving uniform field strength in a region where we have
both a uniform and random field
\begin{equation} \label{ch2:eq:buni}
B_{u} = \left[ \frac{ 4 \pi ( 1 - 2 \alpha) (K_0 + 1) I_\nu
    E_P^{1+2\alpha} (\nu/(2c_1))^{-\alpha}} { (-1 - 2 \alpha)
    c_2(\alpha) \, l \, c_4 \, ((2/3) q^2 + 1)^{(1-\alpha)/2} (q^2
    c_4^{2/(1-\alpha)}+1) } \right] ^ {1/(3-\alpha)}
\end{equation}
In the regions where we have both a random and uniform component to
the field, we calculate $q$ using Equation (\ref{ch2:eq:q}) and then
calculate $B_{u}$ from Equation~(\ref{ch2:eq:buni}) using the total
synchrotron emission and the value of $c_4$ determined from the
polarization angle of the emission using
Equation~(\ref{eq:c4_uni}). The total and random fields can then be
derived from the values of $B_u$, $q$, $c_4$, and $\alpha$.
\begin{eqnarray}
B_{r} & = & q B_{u} c_4^{1/(1-\alpha)} \label{ch2:eq:brelate} \\
B_t & =  & (B_r^2 + B_u^2)^{1/2} \label{ch2:eq:brelate2}
\end{eqnarray}

For the other parameters in Equations~\ref{ch2:eq:buni},
\ref{ch2:eq:brelate}, and \ref{ch2:eq:brelate2}, we determined the
position angle and center of the disk by fitting ellipses to the total
intensity radio continuum image at 3\,cm using the IRAF
\citep{1986SPIE..627..733T,1993ASPC...52..173T} package {\em
  ellipse}. The position angle of the disk is -62.1$^\circ$, which
corresponds to an angle ($\beta$) between the plane of the disk and
the RA axis of 27.9$^\circ$, and the center of the galaxy is
($4^{\rm{h}}30^{\rm{m}}49 \fs 25, 64^\circ 50\arcmin 53 \farcs
0$). The diameter of the disk of NGC 1569 was estimated from the 3\,cm
image to be 2.1~kpc. We assume that the distance and inclination of
the galaxy are 3.36~Mpc \citep{2008ApJ...686L..79G} and 63$^\circ$
\citep{ch2:2002A&A...392..473S}, respectively. Since even small errors
in the determination of the non-thermal fluxes of individual pixels
will correspond to large errors in $\alpha$,\footnote{Here $\alpha$
  represents the spectral index of the {\em non-thermal} emission.} we
estimate a value for $\alpha$ by calculating $\alpha$ for the total
non-thermal flux at 6\,cm and 3\,cm. This method gives a value of
$\alpha= -0.97$. We take $K_0$ to be 100, which is a typical value in
the solar neighborhood, but which may be as low as 40 in regions of
strong shocks \citep{ch2:2005AN....326..414B}. Combining all these
values, we finally determine $B_{u}$, $B_{r}$, and $B_{t}$ for each
pixel in the 3\,cm maps using Equations~\ref{ch2:eq:buni},
\ref{ch2:eq:brelate}, and \ref{ch2:eq:brelate2}.

\begin{figure}
\centering
\includegraphics[scale=1.0]{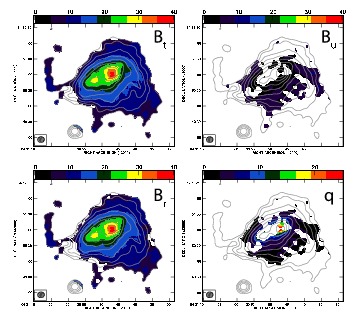}
\caption[Magnetic field strength at 3\,cm for NGC 1569]{ Magnetic
  field strength of NGC 1569 using the 3\,cm data. The color plots
  are, clockwise from upper left, total magnetic field strength (in
  $\mu \rm{G}$), uniform magnetic field strength (in $\mu \rm{G}$),
  $q$, which is the ratio of the random magnetic field strength to the
  uniform magnetic field strength in the plane of the sky (this ratio
  is usually greater than one since the random field is usually much
  stronger than the uniform field), and random magnetic field strength
  (in $\mu \rm{G}$). See \S\ref{ch2:sec:magn-field-prop} for the
  assumptions involved in the derivation of these quantities. The
  total, uniform, and random magnetic field strength have been
  corrected for the assumed geometry of NGC 1569 (see
  Figures~\ref{ch2:fig:ngc1569_geometry_pos} and
  \ref{ch2:fig:ngc1569_geometry_los}).  The 3\,cm total intensity
  emission with $13\arcsec$ resolution is shown by contours. The
  contours are 3, 6, 12, 24, 48, 96, 192, 384, 768, and 1536 times the
  $1\sigma$ noise level.}
  \label{ch2:fig:bfield_3cm}
\end{figure}

Our magnetic maps based on the 3\,cm images with 13\arcsec\ resolution
are shown in Figure~\ref{ch2:fig:bfield_3cm}. The maximum total
magnetic field strength corresponds to the peak in radio continuum
emission at 3\,cm and reaches a value of 38~$\mu \rm{G}$. In the
extended halo of the galaxy, the total magnetic field has a value of
10--15~$\mu \rm{G}$. Examining the plot of $q$ values, i.e., the ratio
of the random field to the uniform field in the plane of the sky, we
see that $q$ is high ($\sim 20$) in the central regions of the galaxy,
suggesting that the random field dominates in these regions, although
there could be significant Faraday depolarization in this region due
to the high electron density. In the outer edges, however, $q$ is only
1--2, so the regular field is quite strong in these regions. This
effect is also seen in the magnetic fields of spiral galaxies
\citep{ch2:1996ASPC...97..457H}. The regular field in the synchrotron
arm ranges between 9 $\mu \rm{G}$ at the southern tip to about 3 $\mu
\rm{G}$ where it connects up with the disk of the galaxy.

To get an idea of how the uncertainties in our parameter estimates
affect our estimated magnetic field strength, we varied each parameter
individually by the estimated error in each parameter, calculated the
difference between the new magnetic field estimate and the nominal
estimate shown in Figure~\ref{ch2:fig:bfield_3cm}, and finally added
the differences from each parameter variation in quadrature. We varied
$\alpha$ by -0.3, $K_0$ to 40 (a value typical for strong shocks), the
distance by 10\%, the diameter by 20\%, the position angle by
3$^\circ$, the center by $6\arcsec$ in each direction, and the
inclination by 3$^\circ$. The resulting calculation showed that the
uncertainties on the total magnetic field strength in the central
region and the outer region of the galaxy were 10~\uG and
3--4~$\rm{\mu G}$, respectively. This error is almost entirely due to
the error in the random field. The error in the determination of the
uniform field is only 0.5--2~\uG. If the \citet{1966MNRAS.133...67B}
formula for $q$ is used instead of the \citet{ch2:1998MNRAS.299..189S}
formula for $q$, the uniform magnetic field is larger by a factor of
1.2 in the center of the galaxy and remains roughly the same at the
edge and the random field remains the same everywhere.

We also investigated how systematic uncertainties in the estimation of
the synchrotron flux due to internal extinction in the central region
of the galaxy will affect the estimated magnetic field. Assuming that
the {\em maximum} internal extinction of 1.74 holds over the entire
galaxy, we get a synchrotron flux down by a factor 0.6 near the main
radio continuum peak, 0.5 near the secondary radio continuum peak, and
0.2 in the region between the two peaks. Since the radio continuum
emission is entirely random in the central region,
\begin{equation}
B_t \propto I_\nu ^{1/(3-\alpha)}
\end{equation}
and using the value of $\alpha$ determined above (-0.97), the maximum
factor by which the magnetic field strength will be decreased is 0.88
for the main radio continuum peak, 0.84 for the secondary radio
continuum beam, and 0.67 for the region between the two peaks.

\subsection{Magnetic Field Structure} \label{ch2:sec:magn-field-struct}

\begin{figure}
\centering
\includegraphics[scale=0.6]{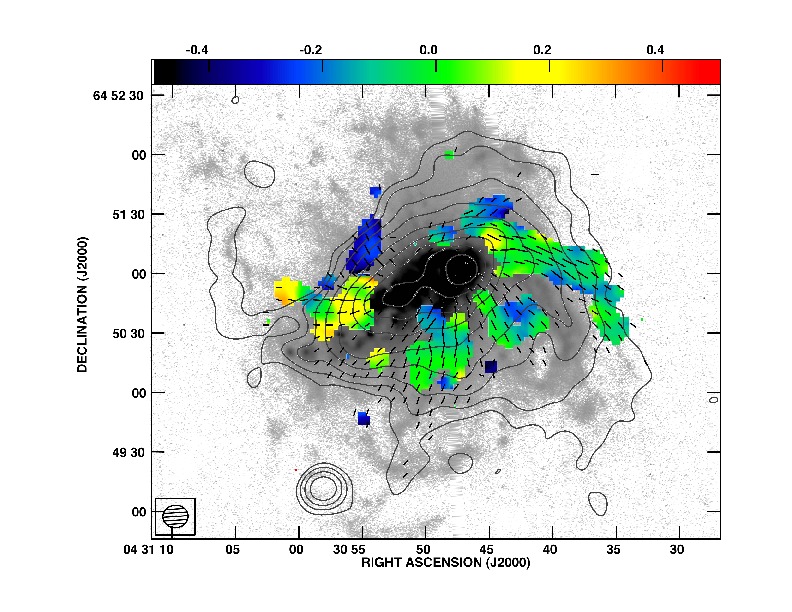}
\caption[Magnetic field structure at 3\,cm of NGC 1569]{ Rotation
  measures between the $14\arcsec$ resolution 6\,cm and 3\,cm data
  (color), magnetic field vectors derived from the 3\,cm data, and the
  contours of the $14\arcsec$ resolution 3\,cm data overlaid on \ha
  image courtesy of D. Hunter \citep{ch2:hunter2004}. The color bar is
  in units of $10^3\ \rm{rad} \ \rm{m}^{-2}$, a 1$\arcsec$
  polarization vector has a polarized intensity of 12.8~\uJybeam, and
  the contours are 3, 6, 12, 24, 48, 96, 192, 384, 768, and 1536 times
  the $1\sigma$ noise.}
  \label{ch2:fig:rotmeas_6cm_3cm}
\end{figure}

The structure of NGC 1569's magnetic field reveals important clues
about the interaction of the magnetic field with the other components
of the interstellar medium and about the mechanisms at work in the
galaxy that may have amplified the field. From our data set, we have
two complementary measures of the structure of the magnetic field.  In
Figure~\ref{ch2:fig:rotmeas_6cm_3cm}, the 3\,cm polarization vectors
rotated by 90$^\circ$ give an estimate of the orientation, but not
direction, of the magnetic field in the plane of the sky. There is
very little Faraday rotation at 3\,cm, so the estimate of the magnetic
field orientation at 3\,cm should be accurate. The difference in the
position angle of the polarized emission at different wavelengths
gives the rotation measure
\begin{equation}
\Delta \phi = {\rm RM} \ \lambda^2
\end{equation}
which depends on the thermal electron density and the magnetic field
along the line of sight
\begin{equation}
{ \rm RM} = 0.812 \int B_\| \, n_e \ {\rm d}s,
\end{equation}
where \rm{RM} is in $\rm{rad} \ m^{-2}$, $n_e$ is in $\rm{cm}^{-3}$,
$B_\|$ is in \uG, and ${\rm d}s$ is in $\rm{pc}$
\citep{1966ARA&A...4..245G}.  The sign of the rotation measure gives
the {\em direction}, not just the orientation of $B_\|$. Positive
rotation measures point towards the observer and negative away. In Figure~\ref{ch2:fig:rotmeas_6cm_3cm}, we plot the rotation measures
between the 6\,cm and 3\,cm data with the estimated magnetic field
vectors from the 3\,cm data overlaid. We do not show the rotation
measures between between the 20\,cm and 13\,cm data because there are
only two patches of significant rotation measures. These patches are
the size of the beam and are located at $4^{\rm{h}} 30^{\rm{m}} 41 \fs
41, 64^\circ 52\arcmin 12 \farcs 98$ (-36 $\rm{rad} \ \rm{m}^{-2}$)
and $4^{\rm{h}} 30^{\rm{m}} 45 \fs 17, 65^\circ 51\arcmin 52 \farcs 0$
(42 $\rm{rad} \ \rm{m}^{-2}$).
The rotation measures were determined using the AIPS task COMB.


Generally, polarization observations at two frequencies do not have
enough information to accurately determine rotation measures
\citep{ch2:1996A&A...313..768R}. However, examining
Figures~\ref{ch2:fig:polarized_intensity_20cm_13cm} and
\ref{ch2:fig:polarized_intensity_3cm_6cm}, we find that the changes in
angle within each pair of frequencies are small enough that we are
able to determine the rotation measure between the frequencies in each
pair with reasonable accuracy.  For the polarization vectors to have
rotated at least 180$^\circ$ between 6\,cm and 3\,cm, the rotation
measure would have to be greater than 1230~$\rm{rad} \ \rm{m}^{-2}$,
which is much larger than the rotation measures seen in other
starbursting galaxies (see \S\ref{sec:comp-magn-field}). Large changes
in the distribution of the polarized emission between 3\,cm and 20\,cm
that prevent the determination of rotation measures using at least
three frequencies are a common feature of magnetic field observations
of edge-on starbursting galaxies
\citep{ch2:1994A&A...282..724R,ch2:1994A&A...292..409B,ch2:1997A&A...320..731D,ch2:1994A&A...284..777G}.

From Figure \ref{ch2:fig:rotmeas_6cm_3cm}, one can see from the
magnetic field vectors that, in general, the magnetic field in the
plane of the sky is roughly perpendicular to the disk of the
galaxy. The magnitude of the rotation measures in the northeastern
half of the galaxy are larger than those in the southwestern half of
the galaxy. The relative magnitude of the rotation measures between
these two regions supports the geometry derived from X-ray
measurements: the southern half of NGC 1569 is inclined towards us,
while the northern half is inclined away
\citep{ch2:2002ApJ...574..663M}. Therefore, the emission from the
region north of the disk passes through a larger column of ionized gas
and thus has a correspondingly larger rotation measure. In the
southeastern portion of the galaxy, there are more rotation measures
close to zero which means that, in general, the field we are seeing is
mostly in the the plane of the sky and that the field along the line
of sight is small.  This overall magnetic field configuration means
that the field can help transfer energy and momentum from cosmic rays
to the thermal plasma
\citep{ch2:1991A&A...245...79B,ch2:1993A&A...269...54B,ch2:2008ApJ...674..258E}
helping to sustain the outflow of gas from NGC 1569.

The magnetic field structure of NGC 1569 is closely tied to the
bubbles of ionized gas seen in \ha (see
Figure~\ref{bubble_finding_chart} for a finding chart of the \ha
bubbles identified in NGC 1569). In Figure
\ref{ch2:fig:rotmeas_6cm_3cm} in the northeastern portion of the
galaxy, we see a pair of oppositely directed rotation measures. The
field along the line of sight in the easternmost region is directed
toward the observer (yellow/green region in the northeastern part),
while the field to the northwest of it is directed away from the
observer (the blue region in the northeastern part). These rotation
measures coincide with \ha bubble E. The combination of the
orientation of the magnetic field vectors in the plane of the sky with
the observed rotation measures suggests a magnetic loop being drawn
out of the disk. The magnetic field vectors at 20\,cm (see
Figure~\ref{ch2:fig:polarized_intensity_20cm_13cm} for the
polarization vectors, which can be rotated by 90$^\circ$ to get an
estimate of the magnetic field vectors) and the two small regions of
significant rotation measures between the 20\,cm data and the 13\,cm
mentioned earlier are associated with \ha Bubble B. The sign of the
rotation measures between the 20\,cm and 13\,cm data is oppositely
directed in the two regions, suggesting another field loop especially
given the estimated orientation of the magnetic field in the plane of
the sky.

Looking closely at the rotation measures between 6\,cm and 3\,cm in
the southern portion of the galaxy, one can see indications of other
possible magnetic field loops associated with \ha bubbles A and F. The
southern patch of rotation measures at
$4^{\rm{h}}30^{\rm{m}}50^{\rm{s}}$ shades from blue to green, which
means that the field along the line of sight is going from being
pointed away from the observer to pointed toward the observer.  This
region is associated with Bubble F. The patch of rotation measures at
$4^{\rm{h}}30^{\rm{m}}42^{\rm{s}}$ shows the opposite trend. The field
goes from being almost completely in the plane of the sky (green) to
pointed away from the observer (blue).  The western arm goes from
yellow where it connects to the disk of the galaxy (field pointed
towards observer) to green (field primarily in the plane of the sky)
to blue-green (field pointed away from observer). These regions are
both associated with Bubble A. 


The bases of all the bubbles are around 30\arcsec\, which at the
distance of NGC 1569 is a physical size of about 500~pc. This size is
roughly double the size of the based of an average superbubble in the
Milky Way (200~pc; \citealp{2001RvMP...73.1031F}).


\subsection{Implications of Derived Magnetic Field Strength and
  Structure} \label{sec:impl-deriv-magn}

\subsubsection{Role of the Field in NGC 1569} \label{sec:role-field-ngc}

From our magnetic field estimates, we can calculate a typical pressure
for the central region and halo of NGC 1569. Adopting 38~$\mu \rm{G}$
for the central field and 12~$\mu \rm{G}$ for the halo field (see
Figure~\ref{ch2:fig:bfield_3cm}), we get pressures ($P/k$) of $4.2
\times 10^5 \ \rm{K \ cm^{-3}}$ and $0.41 \times 10^{5} \ \rm{K \
  cm^{-3}}$ for the central magnetic pressure and the halo magnetic
pressure. In Table~\ref{tab:ism_pressures}, we compare these values
with the pressures determined for other components of NGC 1569's ISM
and the overall gravitational pressure. Where necessary, we have
corrected the input values for the updated distance to NGC
1569. Except for the pressure of the diffuse ionized medium (DIG), we
have not included any corrections for the highly uncertain filling
factors of various components.\footnote{Decreasing the filling factors
  from one, which is what we assume here, will increase the pressure
  by a factor of $1/\sqrt{f}$.}  Note that we associate the broad
component of the turbulent pressure with the DIG and the narrow
component with the HII regions.  Table~\ref{tab:ism_pressures} shows
that the magnetic pressure in NGC 1569 is generally the same order of
magnitude as most of the other sources of pressure in its ISM, but is
also generally less than the other pressures in similar regions. The
dominant pressure source in NGC 1569 is actually the turbulent
pressure of the HII shells, which is not surprising given the level of
disruption in this galaxy. The central x-ray pressure is almost as
large as the gravitational pressure which explains the extended halo
of hot gas around NGC 1569. The pressure balance in NGC 1569 is very
different from the pressure balances in NGC 6946
\citep{2007A&A...470..539B} and M33 \citep{2008A&A...490.1005T}. For
these galaxies, the energy densities (i.e. pressures) of the magnetic
field and the turbulent component are roughly comparable, while the
pressure for the thermal gas is down by a factor of approximately
5. NGC 6946 and M33, however, are very different galaxies from NGC
1569. They are relatively quiescent, star-forming galaxies, unlike NGC
1569, which is in final throes of a starburst that has significantly
affected its ISM.





\subsubsection{Generating the Field} \label{sec:generating-field}

\citet[][and references therein]{ch2:2000A&A...358..125F} and
\citet{2008A&A...486L..35G} have developed models in which galactic
rotation provides the shear and supernova explosions drive the
turbulence needed to generate a large scale magnetic field. In these
models, one would expect the magnetic field to be correlated with the
supernovae driven bubbles as seen in NGC 1569 (see
\S~\ref{ch2:sec:magn-field-prop}). Note that this is a necessary, but
not sufficient condition for a supernova driven dynamo to work. The
calculations of Ferri{\`e}re et al.\ and Gressel et al.\ are based on
mean field $\alpha - \omega$ dynamo models, which have been criticized
on theoretical grounds
\citep{2008RPPh...71d6901K,2009MNRAS.395L..48C}. Nevertheless, whether
or not mean field theory is valid, it seems likely that shear and
turbulence are key ingredients in a galactic dynamo.  Therefore, we
estimate the strength of these two components (turbulence via
supernova explosions and shear via differential rotation) for NGC
1569.

First, we compare the supernova rate in the Milky Way to that of
NGC~1569. The supernova rate from type II supernovae in the Milky Way
is 27~$\rm{Myr}^{-1}\ \rm{kpc}^{-2}$
\citep{2001RvMP...73.1031F}.\footnote{Including type I supernovae
  increases the total rate by 15-20\%.}
\citet{ch2:2005AJ....129.2203A} find that the recent star formation
rate for NGC 1569 is $6.4 \ \rm{M_\sun \ yr^{-1} \ kpc^{-2}}$. This
star formation rate was determined using a distance estimate of
2.9~Mpc. Increasing the distance used in the star formation rate
estimate to the best current estimate (3.36~Mpc) will increase the
rate by a factor of two to three and cause the latest epoch of star
formation to end recently \citep{2008ApJ...686L..79G}. Assuming a
Salpeter IMF with 0.8 and 150~$\rm{M}_\sun$ as the minimum and maximum
stellar masses and normalizing it to the total number of stars formed
per year and per $\rm{kpc}^{2}$, we find that that the number of stars
formed with masses greater than 8~$\rm{M_\sun}$ for NGC 1569 is $1.1
\times 10^5 \ \rm{Myr^{-1}} \ \rm{kpc^{-2}}$. With a duration of the
most recent burst of star formation of 19~Myr
\citep{ch2:2005AJ....129.2203A} and a typical lifetime for an
8~$\rm{M_\sun}$ star of 30~Myr, we get a supernova rate of $7 \times
10^4 \ \rm{Myr^{-1}} \ \rm{kpc^{-2}}$.  This rate is a factor of 2500
higher than the supernova rate in the Milky Way.

The shear is defined as 
\begin{equation}
\frac{\rm{d} \, \Omega}{\rm{d} \, R}
\end{equation}
where $\Omega$ is the angular speed ($v/R$) and $R$ is the
radius. Given that the rotation velocity in the outer regions of
galaxy disks is roughly constant, the shear can be estimated by
\begin{equation}
  \left| \frac{\rm{d} \, (\it{v} /R)}{\rm{d} \, R} \right| = \frac{v}{R^2}.
\end{equation}
In the Milky Way, the angular speed is approximately 220~\kms at a
radius of 8.0~kpc, so the shear is 3.4~$\rm{km \, s^{-1}} \,
\rm{kpc^{-2}}$. In NGC 1569, the galaxy has a rotation speed
(corrected for an inclination of 60$^\circ$) of 46~\kms at a radius of
1.4~kpc \citep{ch2:2005AJ....130..524M}, which gives a shear of 33
$\rm{km \, s^{-1}} \, \rm{kpc^{-2}}$, which is a factor of 10 larger
than the shear in the Milky Way.

Since both the shear and the supernovae rate are much greater than in
the Milky Way, the fidicual model for this type of dynamo, it is quite
possible for a supernova-driven dynamo like those described in
\citet{ch2:2000A&A...358..125F} and \citet{2008A&A...486L..35G} to be
at work in NGC 1569. Detailed modeling, however, is needed to
determine whether or not this type of dynamo is at work in NGC
1569. Note that the supernova-driven dynamo described above is
different that the related mechanism modeled described by
\citet{1992ApJ...401..137P} and modeled by
\citet{2009ApJ...693....1O}. In that mechanism, the magnetic buoyancy
driven by cosmic rays lofts the field loops into the halo of the
galaxy.

\subsubsection{Comparison to Magnetic Fields of Other Starbursts} \label{sec:comp-magn-field} 

Table~\ref{tab:bfield_strengths_starbursts} summarizes the estimated
magnetic field strengths in other starburst galaxies. Overall, we find
that our estimate for the magnetic field strength of NGC 1569 is in
the middle of the distribution of magnetic field strengths for
starbursting galaxies with NGC 1569's field being greater than that of
NGC 253 and NGC 4666, but less than that of M82 and NGC 1808. It is
important to note, however, that these galaxies are all much larger
than NGC 1569, so for the field in NGC 1569 to be even comparable to
that in larger galaxies has interesting implications for dynamo
theory. \citet{2006ApJ...645..186T} argue that the magnetic field
strengths of starbursts like NGC 253, M82, and NGC 1808 are
underestimated by the minimum energy magnetic field estimates based on
synchrotron emission and argue for much greater fields strengths in
starbursting galaxies based on equipartition of the gravitational and
magnetic pressures. However, observations of Zeeman splitting in
(U)LIRGS by \citet{2008ApJ...680..981R}, which yield direct
measurements of the magnetic field strength, agree roughly with the
values derived from synchrotron emission.


Comparing the magnetic field structure we derive for NGC 1569 (see
Figure~\ref{ch2:fig:bfield_3cm}) with the magnetic fields structures
derived for other starburst galaxies (NGC 253, M82, NGC 1808, NGC
4569, and NGC 4446), we can see several important differences.  First,
the polarization features in NGC 1569 are patchier than in any of the
other galaxies. There are two probable causes for this patchiness: 1)
the magnetic field of NGC 1569 is much more turbulent and 2) the
depolarization effects in NGC 1569 are much more severe. Given the
turbulent nature of the \ha in the central regions of NGC 1569
\citep{ch2:2008MNRAS.383..864W}, it is likely that the field is very
turbulent in this region as well, although the large electron density
here makes it possible for Faraday depolarization to mask any uniform
field component. However, in the outer regions of this galaxy
depolarization effects due to the \ha filaments are likely to dominate
over turbulence given the organization imposed on the ISM by the gas
outflow. In addition, if the depolarization in the outer regions was
due to turbulence it would mean that the turbulence would have to vary
in the halo of the galaxy, which is at odds with the result of
\citet{ch2:2008MNRAS.383..864W}, who found that the warm ionized gas
in the outer regions has more or less constant velocity width between
20 to 70 \kms. The second difference between the previous observations
and our observations of NGC 1569 is that the ratio of the length of
the polarized arm to the major axis of the radio continuum emission
associated with the galactic disk is very high ($\sim 0.5$). Of the
previous observations of starbursting galaxies, only NGC 4569
\citep{ch2:2006A&A...447..465C}, whose interstellar medium has been
shaped by its location in the Virgo cluster, has such a large ratio of
radio continuum feature length to length of the major axis of the
radio continuum emission associated with the disk.

Rotation measures have been determined for M82
\citep{ch2:1994A&A...282..724R}, NGC 253
\citep{ch2:1994A&A...292..409B}, and NGC 4446
\citep{ch2:1997A&A...320..731D}. Comparing our derived rotation
measures in Figure~\ref{ch2:fig:rotmeas_6cm_3cm} to those derived
between 6\,cm and 3\,cm for M82 and NGC 253, we see that the average
rotation measure between these two frequencies is more or less the
same for all three galaxies ($\sim \pm 200 \ \rm{rad} \,
\rm{m^{-2}}$). The rotation measures for NGC 4666 were determined
between 20\,cm and 6\,cm and their magnitudes correspond well to the
rotation measures determined here between 20\,cm and 13\,cm ($\sim \pm
40 \ \rm{rad} \, \rm{m^{-2}}$).



\section{A Multiwavelength Picture of the Interstellar Medium of NGC
  1569} \label{ch2:sec:discussion}

NGC 1569 is a well-studied galaxy. In this section, we compare our
observations with those at other wavelengths to examine the larger
picture of the complex interstellar medium of NGC 1569. We start with
the \ha emission from NGC 1569, which is characterized by long \ha
filaments extending along the minor axis of the galaxy. The filaments
were first discussed extensively by \citet{ch2:1974ApJ...191L..21H}
and were the subject of much subsequent work
(e.g. \citealt{ch2:1974ApJ...194L.119D,ch2:1991ApJ...370..144W,ch2:1993AJ....106.1797H,ch2:1995ApJ...448...98H,ch2:1997ApJ...475...65H}). \citet{ch2:2007MNRAS.381..894W,ch2:2007MNRAS.381..913W}
show that in the central region the emission lines have two
components: a narrow-line component ($\rm{FWHM} \le 70 \ \rm{km \
  s^{-1}}$) that they associate with the turbulent ISM and a
broad-line component ($\rm{FWHM} \sim 150 \ \rm{km \ s^{-1}}$) that
they associate with emission from a turbulent mixing layer between
cool gas clumps and the hot gas from surrounding star clusters. The
broad-line component of the emission is confined to the central region
of the galaxy \citep{ch2:2008MNRAS.383..864W}. The outer regions of
NGC 1569 show clear signs of an outflow of gas from the center
\citep{ch2:2008MNRAS.383..864W}. This result agrees with previous
results from long-slit spectra
\citep{ch2:1994PASJ...46..335T,ch2:1995ApJ...448...98H,ch2:1998ApJ...506..222M}.

Figures~\ref{ch2:fig:polarized_intensity_20cm_13cm} and
\ref{ch2:fig:polarized_intensity_3cm_6cm} compare the distribution of
the \ha emission to the radio continuum emission and were partially
discussed in \S~\ref{ch2:sec:polarized-emission}
and~\ref{ch2:sec:magn-field-struct}. Here we discuss three additional
points. First, the radio continuum arm becomes increasingly aligned
with the \ha arm as the wavelength decreases. This fits well with the
region on the outside of the \ha arm being synchrotron dominated,
which is what one would expect from our thermal fraction estimates in
\S~\ref{ch2:sec:estim-therm-flux}.  Second, the uniform component of
the magnetic field in this region may be at least partially due to
compression by the expanding \ha bubble in that region. In
Figure~\ref{ch2:fig:bfield_3cm}, the region of uniform magnetic field
lies outside of (to the west) of the \ha arm. Finally, the presence of
a magnetic field in the \ha arm region may explain why the western \ha
arm is thicker than the other \ha features in this galaxy (see
Figures~\ref{ch2:fig:polarized_intensity_20cm_13cm} and
\ref{ch2:fig:polarized_intensity_3cm_6cm}). The field here could
provide an additional source of pressure, working against any
compression forces, thereby increasing the width of the filament
compared to the filaments elsewhere in the galaxy. However, the effect
of the magnetic field on the arm is depends on a variety of factors
including the magnetic field strength in the arm, the variation of
thermal pressure with height above the galaxy disk, and the effect of
the starburst on the distribution of ISM of NGC 1569. In addition, our
data only gives the integrated magnetic field -- we cannot determine
what the field is in the \ha arm, only the integrated field in that
region, which is dominated by depolarized effects. Detailed modeling
is the only way to disentangle these effects.

\begin{figure}
\centering
\includegraphics{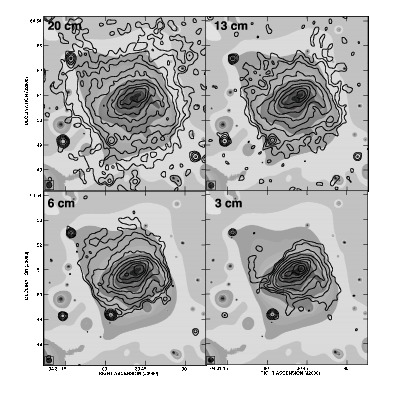}
\caption[Radio continuum emission compared to X-ray emission for NGC
1569]{ Adaptively smoothed, total ($0.3 \ \rm{keV} \leq \rm{E} \leq
  8.0 \ \rm{keV}$) X-ray emission image from
  \citet{ch2:2005MNRAS.358.1423O} with 20\,cm, 13\,cm, 6\,cm, and
  3\,cm contours overlaid. The radio continuum images all have beams
  with approximately the same size as the 20\,cm beam. The contours
  start at 3 times the $1\sigma$ noise level in each band and each
  subsequent contour increases by a factor of 2. The beam for each
  wavelength is in the lower left of each panel. The region of dark
  gray in the lower left hand corner of each panel is not emission
  from NGC 1569 but instead is due to edge effects from the adaptive
  smoothing process.} \label{ch2:fig:total_intensity_xray}
\end{figure}

X-ray observations of NGC 1569 reveal that the galaxy is surrounded by
a halo of hot gas bounded by \ha filaments
\citep{ch2:1995ApJ...448...98H,ch2:1996ApJ...469..662D}. High spatial
resolution observations by \citet{ch2:2002ApJ...574..663M} have shown
that the bright X-ray emission near the filaments is most likely due
to shocks generated by the outflow of gas from NGC 1569. The measured
ratio of $\alpha$ elements to iron is between 2.5 and 4.8 times the
solar ratio, indicating that the wind is laden with the products of
type II supernovae \citep{ch2:2002ApJ...574..663M}. We compare our
radio continuum observations with the adaptively smoothed X-ray map of
\citet{ch2:2005MNRAS.358.1423O} in
Figure~\ref{ch2:fig:total_intensity_xray}. To compare the extents of
the X-ray and radio continuum maps, we have derived scale lengths for
the radio continuum emission by azimuthally averaging the emission as
a function of radius from the center of the emission using IRING in
AIPS. We find that the scale lengths are 376.1~pc (20\,cm), 310.9~pc
(13\,cm), 236.5~pc (6\,cm), and 258.7~pc
(3\,cm). \citet{ch2:2005MNRAS.358.1453O} found that the scale length
for the X-ray emission is 208 pc, so the radio continuum emission
extends further than the X-ray emission at all wavelengths. We suggest
that the magnetic field in the radio continuum arm may confine the hot
gas on the western edge of the galaxy and funnel the gas toward the
halo of the galaxy. Similar effects are seen in magnetohydrodynamic
simulations of the ISM by \citet{ch2:2005A&A...436..585D}.





\begin{figure}
\centering
\includegraphics{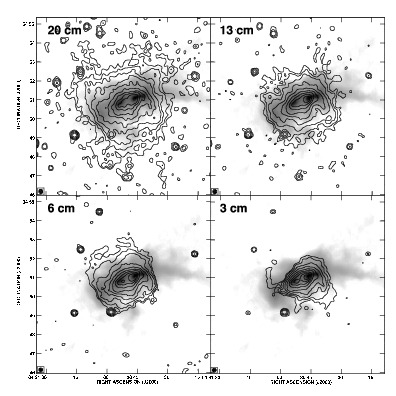}
\caption[Radio continuum emission compared to neutral hydrogen
emission for NGC 1569]{ Neutral hydrogen emission from
  \citet{ch2:2005AJ....130..524M,ch2:2006AJ....132..443M} with 20\,cm,
  13\,cm, 6\,cm, and 3\,cm contours overlaid. The radio continuum
  images all have beams with approximately the same size as the 20\,cm
  beam. The contours start at 3 times the $1\sigma$ noise level and
  increase by a factor of two for each subsequent contour. The beam
  for the radio continuum images is shown the lower left of each
  panel. }
\label{ch2:fig:total_intensity_hi_mom0}
\end{figure}

The neutral hydrogen distribution of NGC 1569 is quite complex
(cf. \citealp{ch2:2002A&A...392..473S,ch2:2005AJ....130..524M}~and
\citealp{ch2:2006AJ....132..443M}). The HI distribution has an arm to
the west of the galaxy, but this feature is distinct from the arm seen
in X-ray, $\rm{H}\alpha$, and radio continuum.
\citet{ch2:2005AJ....130..524M} suggest that the HI arm may be the
remnant of a primordial gas cloud that interacted with NGC 1569 and
started the most recent burst of star formation in this galaxy; the HI
arm is likely still in the process of accreting onto the disk of the
galaxy. In Figure~\ref{ch2:fig:total_intensity_hi_mom0}, we compare
the neutral hydrogen distribution to our radio continuum
observations. The location of the polarized western arm south of the
HI arm also suggests an interaction between the infalling HI arm and
the galaxy itself. We speculate that the strong magnetic field
associated with the \ha arm is the result of the collision of the HI
arm with the expanding bubble (Bubble A in
\citealp{ch2:2008MNRAS.383..864W} and Filament 6 in
\citealp{ch2:1993AJ....106.1797H}). The western \ha arm is
morphologically very different from the other filaments in the galaxy
and given the confluence of interesting $\rm{H}\alpha$, HI, and radio
continuum features in this region, it would be surprising if they were
not related.


The extended radio halo around NGC 1569 may not be unique to this
galaxy. In general, dwarf galaxies have very shallow potential wells,
so any significant star formation episodes will tend out drive
material out of their potential wells. This mechanism has been
postulated as a way to enrich the intergalactic medium
\citep{ch2:1999ApJ...513..142M}, but it could also be a mechanism to
magnetize the intergalactic medium
\citep{ch2:1999ApJ...511...56K,ch2:2000ApJ...541...88V,ch2:2006MNRAS.370..319B}. If
these halos are like the radio continuum halo of NGC 1569, then they
will be dominated by synchrotron emission. Therefore, the best
wavelength regime to search for them is the low frequency radio regime
with telescopes like LOFAR \citep{2008IAUS..255..167K}.  The detection
of such halos in galaxies other than NGC 1569 would provide important
clues about the magnetization of the universe and the escape of cosmic
rays from the halos of galaxies.


\section{Summary and Conclusions} \label{ch2:sec:summary-conclusions}

NGC 1569 provides a crucial test of the importance of magnetic fields
in starbursting dwarf galaxies.  We present the first observations to
date of the polarized radio continuum emission at 20\,cm, 13\,cm,
6\,cm, and 3\,cm from the post-starburst dwarf irregular NGC 1569.  We
detect radio continuum emission from NGC 1569 at all four
wavelengths. Our observations are in agreement with previous
high-resolution radio continuum observations of NGC 1569
\citep{ch2:1976A&A....48..421S,ch2:1983ApJS...53..459C,ch2:1986A&A...161..155K,ch2:1988A&A...198..109I,ch2:2004MNRAS.349.1335L}. We
confirm the discovery of an extended radio continuum halo, which was
first imaged by \citet{ch2:1983ApJS...53..459C} and discussed by
\citet{ch2:1988A&A...198..109I}, and for the first time clearly image
the western arm at wavelengths shorter than 20\,cm. In general, the
spectral indices derived for this galaxy get steeper as one moves
along the minor axis of the galaxy away from the disk. However, there
are two filamentary regions of flat spectral indices extending to the
edge of the map. The spectral index trends shown here agree with those
found by \citet{ch2:2004MNRAS.349.1335L} supporting their theory that
there is a convective wind at work in the halo of this galaxy.

We detect strong polarized emission at all four frequencies. The
polarized emission at 6 and 3\,cm is spread over most of the galaxy,
while the polarized emission at 20\,cm is much spottier, and there are
only a few small patches of polarized emission at 13\,cm.  The
polarized emission at 20\,cm, 6\,cm and 3\,cm is associated with
regions identified as bubbles from \ha observations. There is little
polarization in the central region of NGC 1569, where the ISM is very
turbulent and the largest thermal electron column density can be
found. The \ha filaments in the halo also depolarize the emission.

In order to determine the magnetic field strength, we estimate the
thermal contribution to the radio continuum emission using a
flux-calibrated \ha image. We find that the thermal contribution peaks
at 40--50\% in the center of the galaxy and falls off quickly. The
halo is dominated by synchrotron emission. The total magnetic field
strength in the central regions is 38~\uG and approximately 10--15~\uG
in the halo. The field is largely random near the center; the uniform
field is strongest in the halo.  The magnetic field structure of NGC
1569 is shaped by the outflow of gas from this galaxy drawing field
loops out of the disk. We show that the magnetic field structure of
NGC 1569 is closely aligned with the \ha bubbles.

We find that the magnetic pressure is the same order of magnitude as
other components of the pressure in the ISM, but is generally less
than the pressures of other components in similar regions by a factor
of a few. Therefore, the magnetic field does not dominate the ISM of
NGC 1569, although it clearly plays an important role.

The close correspondence between the \ha bubbles and the magnetic
field structure suggests that the interstellar medium of NGC 1569
contains magnetic bubbles similar to those modelled by
\citet{1991ApJ...375..239F}. The high supernova rate and strong shear
in the galaxy compared to the Milky Way suggests that these bubbles
might combine to give a supernova/superbubble driven dynamo like that
of \citet{ch2:2000A&A...358..125F} and \citet{2008A&A...486L..35G}.
However, detailed modeling is necessary to say definitely whether or
not this type of dynamo is at work in NGC 1569.

Finally, we compare our data to observations of the ISM of NGC 1569 at
other wavelengths. The turbulent \ha kinematics of the central portion
of the galaxy are reflected in a primarily random field, while the
outflow kinematics in the halo are reflected by a uniform field. The
magnetic field in the \ha arm may be increasing the width of this
feature by providing an additional source of pressure as well as
channeling hot gas out of the plane. We suggest that the infall of the
western HI arm into the disk of NGC 1569 and the outward expansion of
the bubble associated with the western \ha arm compressed the ISM in
this region and produced the strong uniform magnetic field seen
here. In short, our observations reveal that the magnetic field of NGC
1569 is shaped by the outflow of gas from the galaxy and its
interaction history and that the magnetic field may be playing an
important role in directing the outflow of gas from the galaxy.

Radio continuum halos generated by the outflow of gas from galaxy may
be common in dwarf starburst galaxies. These halos provide a mechanism
for magnetizing the intergalactic medium
\citep{ch2:1999ApJ...511...56K,ch2:2000ApJ...541...88V,ch2:2006MNRAS.370..319B}. Surveys
of edge-on dwarf starbursts with future low-frequency radio telescopes
like LOFAR will help constrain the number of dwarf starbursts with
synchrotron-dominated radio halos and provide important clues about
the magnetization of the universe and the escape of cosmic rays from
galaxy halos.

\acknowledgments 

The authors would like to thank Deidre Hunter and J{\"u}rgen Ott for
providing us with the \ha and X-ray images of NGC 1569 used in this
paper. We also thank Rainer Beck, Crystal Brogan, Ger de Bruyn,
Krzysztof Ch{\.y}zy, Bill Cotton, Bryan Gaensler, Jay Gallagher,
Miller Goss, Dan McCammon, J{\"u}rgen Ott, and Sne\v{z}ana
Stanimirovi\'{c} for helpful conversations. Finally, we would also
like to acknowledge the extremely helpful comments made by the
anonymous referee. A.A.K.\ was supported by an NSF Graduate
Fellowship, a Wisconsin Space Grant Graduate Fellowship, and a GBT
Student Support award during portions of this work.  S.M.\
acknowledges the Deutsche Forschungsgemeinschaft for the award of a
fellowship of the Graduiertenkolleg ``The Magellanic System, Galaxy
Interaction, and the Evolution of Dwarf Galaxies'', support under
grant SFB~494, and the hospitality of the Department of Astronomy at
University of Wisconsin--Madison for her visit in February
2007. E.G.Z., J.E., and E.M.W. would like to acknowledge the support
of NSF grants AST 0507367 (E.G.Z. and J.E.), PHY 0215581 (E.G.Z. and
J.E.), and AST 0708002 (E.M.W.).


\appendix

\section{Deriving the Geometric Corrections} \label{ch2:sec:deriv-geom-corr}

\begin{figure}
\centering
\includegraphics{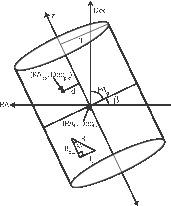}
\caption{ Geometry assumed for NGC 1569 in plane of sky.} \label{ch2:fig:ngc1569_geometry_pos}
\end{figure}

\begin{figure}
\centering
\includegraphics{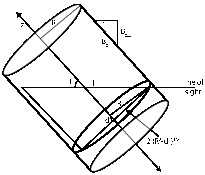}
\caption{ Geometry assumed for NGC 1569 along line of sight} \label{ch2:fig:ngc1569_geometry_los}
\end{figure}

We assume that the geometry of NGC 1569 can be approximated by an
infinite cylinder with the vector normal to the galaxy disk inclined
from the line of sight by $i$ degrees ($i=0^\circ$ is face-on,
$i=90^\circ$ is edge-on) and rotated on the sky by a position angle
(measured from north and increasing counterclockwise, i.e, towards the
east) $\rm{PA}_g$. The geometry for NGC 1569 and the relevant
variables are shown in Figures~\ref{ch2:fig:ngc1569_geometry_pos} and
\ref{ch2:fig:ngc1569_geometry_los}. Below we derive the relevant
quantities needed for our calculation of the magnetic field strength.

\subsection{Deriving the Line of Sight
  Distance}\label{ch2:sec:deriving-line-sight}

The line-of-sight length ($l$) through an inclined (infinite) cylinder
is
\begin{equation} \label{ch2:eq:los_through_cylinder}
l = \frac{2 \ (R^2-d^2)^{1/2} } { \sin{i}}
\end{equation}
where $d$ is the distance from the central axis of the cylinder, $R$
is the radius of the cylinder, and $i$ is the inclination from the
line of sight. If the cylinder is rotated about its minor axis in the
plane of the sky, where the angle between the RA axis and the disk of
the galaxy (minor axis of cylinder) is $\beta$, $d$ becomes
\begin{equation}\label{ch2:eq:d_wrt_to_rotation}
d = D 
\frac{\pi}{180} 
\left[  ( \rm{RA_{pix}} - \rm{RA_0}) \cos{(\rm{Dec_0})} \cos{\beta} 
- (\rm{Dec_{pix}} - \rm{Dec_0}) \sin{\beta} \right]
\end{equation}
where $D$ is the distance to the galaxy,
$(\rm{RA_{pix}},\rm{Dec_{pix}})$ are the coordinates of a particular
pixel, and $(\rm{RA_{0}},\rm{Dec_{0}})$ are the coordinates of the
center of the galaxy. The variable $\beta$ is defined as positive for
counterclockwise rotations and negative for clockwise rotations and is
$90 + \rm{PA}_g$ for $\rm{PA}_g < 0$ and $\rm{PA}_g-90$ for $\rm{PA}_g
> 0$. If $\beta$ is 0$^\circ$, then $d$ is just
\begin{equation}
d = D \frac{\pi}{180} (\rm{RA}_{pix} - \rm{RA}_0) \cos{(\rm{Dec}_0)}
\end{equation}

\subsection{Deriving $c_4$ for our geometry.} \label{ch2:sec:deriving-c_4}

To deal with the position angle of the galaxy, we correct the observed
polarization vector position angles ($\rm{PA}_{B,O}$), which are on
the usual position angle system where $\rm{PA}_g=0\arcdeg$ is north,
$\rm{PA}_g=90\arcdeg$ is east, and $\rm{PA}_g=-90\arcdeg$ is west, so
that they are centered on a coordinate system where a position angle
of $0\arcdeg$ is along the axis of the cylinder and a position angle
of $90\arcdeg$ is pointing toward the left (eastward) and a position
angle of $-90\arcdeg$ is pointing toward the right (westward).  For a
negative position angle, the correction is
\begin{equation}
\rm{PA}_{B} = -(90\arcdeg+\rm{PA}_g) + \rm{PA}_{B,O}
\end{equation} 
and for a positive position angle
\begin{equation}
\rm{PA}_{B} = (90\arcdeg-\rm{PA}_g) + \rm{PA}_{B,O}
\end{equation}
We rotate any resulting $\rm{PA}_{B}$ outside the range of [90,-90] by
$180\arcdeg$.

In the frame of the cylinder, there are two components to the observed
magnetic field vector ($B_{\perp}$): $B_x$ and $B_{z,\perp}$. The
$B_x$ components are not affected by the inclination of the galaxy
because they are perpendicular to the inclination axis. In the $z$
direction, however, we only see the field strength $B_z$ projected
onto the plane of the sky ($B_{z,\perp}$). The observed magnetic field
strength in the $z$ direction is
\begin{equation}
B_{z,\perp} = B_z \sin{i}
\end{equation}
where $B_z$ is the magnetic field strength in the $z$ direction and
$i$ is the inclination. Now the observed magnetic field strength
($B_\perp$) is also equal to
\begin{equation}
B_{z,\perp} = B_{\perp} \cos{(\rm{PA}_B)}
\end{equation}
Therefore,
\begin{equation}
B_z = B_{\perp} \cos{(\rm{PA}_B)} / \sin{i}.
\end{equation}
Finally, the total magnetic field ($B_t$) is
\begin{eqnarray}
B_t &  = & ( B_x^2 + B_z^2 ) ^{1/2} \\
    &  =  &\left[ ( B_{\perp} \sin{(\rm{PA}_B)} )^2 + \left( \frac{B_{\perp}
    \cos{(\rm{PA}_B)} } {\sin{i}}\right)^2 \right] ^ {1/2} \\
   & = &   B_{\perp} \left[ (\sin{(\rm{PA}_B)})^2 + \left(\frac{ \cos{(\rm{PA}_B)}} {  \sin{i}}\right)^2 \right]^{1/2}
\end{eqnarray}
\citet{ch2:2005AN....326..414B} define $c_4$ as
\begin{equation}
c_4 = \left(\frac{B_\perp}{B_t}\right)^{(1-\alpha)}
\end{equation}
so our $c_4$ is then
\begin{equation}
c_4 =  \left[ (\sin{(\rm{PA}_B)})^2 + \left(\frac{ \cos{(\rm{PA}_B)} }  {
    \sin{i}}\right)^2 \right] ^ {- (1 - \alpha ) / 2}
\end{equation}
for a completely regular field. 


\bibliography{n1569}


\begin{deluxetable}{rcc}
\tablewidth{0pt}
\tabletypesize{\scriptsize}
\tablecaption{Summary of WSRT 20\,cm Observations \label{tab:wsrt_20cm_obs_summary}}
\tablehead{ 
  \colhead{} &
  \colhead{ 5--6 Nov}  &
    \colhead{ 6--7 Nov } \\
  \colhead{}     & 
  \colhead{2000} &
  \colhead{2000}  } 
\startdata
Array Configuration  &  36m         &   72m               \\
IF1 (GHz)	     & 1.345        &  1.345                 \\
IF2 (GHz)            & 1.355        &  1.355                \\
IF3 (GHz)            & 1.365        &  1.365                \\
IF4 (GHz)            & 1.375        &  1.375                \\
IF5 (GHz)            & 1.385        &  1.385                \\
IF6 (GHz)            & 1.395        &  1.395                \\
IF7 (GHz)            & 1.405        &  1.405                \\
IF8 (GHz)            & 1.415         &  1.415                \\
Bandwidth per IF (MHZ)    & 10           &  10               \\
Flux Calibrator      & 0137+331     & 1331+305          \\ 
Instrumental         & 0137+331     & 0542+498          \\
Polarization         &              &                   \\
Absolute             &  2253+161    & 1331+305          \\
Polarization Angle   &              &                   \\
\enddata
\end{deluxetable}

\begin{deluxetable}{cccccccccc}
\tablewidth{0pt}
\tabletypesize{\scriptsize}
\tablecaption{Final Images \label{tab:final_image_summary}} 
\tablehead{ 
  \colhead{ } &
  \colhead{Frequency} &
  \colhead{} &
  \colhead{Beam} &
  \colhead{PA} &
  \colhead{LAS\tablenotemark{a}} &
  \colhead{$\sigma_I$}  &
  \colhead{$\sigma_Q$} &
  \colhead{$\sigma_U$} \\
  \colhead{Data Name} &
  \colhead{GHz} &
  \colhead{Weighting}     & 
  \colhead{$\arcsec$} &
  \colhead{$^\circ$} &
  \colhead{\arcmin} &
  \colhead{$\mu$Jy beam$^{-1}$} &
  \colhead{$\mu$Jy beam$^{-1}$} &
  \colhead{$\mu$Jy beam$^{-1}$} 
} 
\startdata
20cm            & 1.38  & Natural\tablenotemark{b}   & $12.88 \times 14.19$   &  6.5    & 21                       & 16.5 & 12.6  &  13.2    \\    
\hline
13\,cm, robust=0  & 2.27  & Robust = 0                 & $10.19 \times 9.49$    &  -9.46  & 13                       & 19.6 & 17.2  & 17.3    \\
13\,cm, 13\arcsec & 2.27  & uvrange=(0.165,13)         & $13.44 \times  12.96$  &  -9.60  & 13                       & 24.8 & 19.4  & 19.3    \\
\hline
6\,cm, robust=0  & 4.86   & Robust = 0                 & $4.42 \times   4.08$  &  89.77   & \nodata\tablenotemark{c} & 8.5  & 7.6   & 7.6   \\ 
6\,cm, 13\arcsec & 4.86   & uvrange=(0.165,13)         & $12.64 \times  11.66$ &  -88.20  & \nodata\tablenotemark{c} & 14.0 & 10.8  & 10.7  \\
\hline
3\,cm, robust=0 & 8.46    & Robust = 0                 & $8.85 \times   6.82$  & -78.31   & \nodata\tablenotemark{c} & 11.1   & 8.0   &  8.0 \\
3\,cm, 13\arcsec& 8.46    & uvrange=(0.165,13)         & $12.65 \times  11.26$ & -86.44   & \nodata\tablenotemark{c} & 17.6  & 10.8  &  11.0 \\
\enddata
\tablenotetext{a}{ {Largest Angular Scale}}
\tablenotetext{b}{ All fixed-moveable baselines selected.}
\tablenotetext{c}{ This image is the result of combining VLA and Effelsberg observations, so it includes all spatial frequencies greater than the frequency of the synthesized beam. }
\end{deluxetable}

\begin{deluxetable}{rcc}
\tablewidth{0pt}
\tabletypesize{\scriptsize}
\tablecaption{Summary of WSRT 13\,cm Observations \label{tab:wsrt_13cm_obs_summary}}
\tablehead{ 
  \colhead{} &
  \colhead{ 4-5 Dec } &
  \colhead{ 18-19 Jan  } \\
  \colhead{}     & 
  \colhead{2002} &
  \colhead{2003} }
\startdata
Array Configuration &  72m        &   36m       \\ 
IF1 (GHz)	    & 2.200000    &  2.20000    \\ 
IF2 (GHz)	    & 2.218125    & 2.218125    \\ 
IF3 (GHz)           & 2.236250    & 2.236250    \\ 
IF4 (GHz)           & 2.254375    & 2.254375    \\ 
IF5 (GHz)           & 2.272500    & 2.272500    \\ 
IF6 (GHz)           & 2.290625    & 2.290625    \\ 
IF7 (GHz)           & 2.308750    & 2.308750    \\ 
IF8 (GHZ)           & 2.326875    & 2.326875    \\ 
Bandwidth per IF (MHz) &  20         & 20          \\ 
Bandpass Calibrator & 1331+305    & 1331+305    \\ 
Flux Calibrator     & 1331+305    & 1331+305    \\ 
Instrumental        & 0834+555    & 0834+555    \\ 
Polarization        &             &             \\ 
Absolute            & 1331+305    &  1331+305   \\ 
Polarization Angle  &             &             \\ 
R-L Angle           &  66         &  66         \\ 
\enddata
\end{deluxetable}

\begin{deluxetable}{rcccccccc}
\tablewidth{0pt}
\tabletypesize{\scriptsize}
\tablecaption{Summary of VLA  6\,cm and 3\,cm Observations \label{tab:vla_obs_summary}}
\tablehead{ 
  \colhead{} &
  \colhead{ 11 Aug. } &
  \colhead{ 13 Aug.  } &
  \colhead{ 17 Aug.  } &
  \colhead{ 27 Aug.  } &
  \colhead{ 19 Sept.  } &
  \colhead{ 23 Sept.  } &
  \colhead{ 9 Aug.  } &
  \colhead{ 10 Aug. } \\
  \colhead{}     & 
  \colhead{2000} &
  \colhead{2000} &
  \colhead{2000} &
  \colhead{2000} &
  \colhead{2000}  &
  \colhead{2000} &
  \colhead{2001} &
  \colhead{2001}
}
\startdata
Program Number   & AM643       & AM643       & AM643        & AM643       & AM643       & AM643       & AM694         & AM694          \\
Array            &  D          &   D         &   D          &   D         &   D         &   D         &   C           &   C            \\
IF1 (GHz)	 &  8.4351     & 4.8851      & 8.4351       & 8.4351      & 8.4351      & 8.4351      & 4.8851        & 4.8851         \\
IF2 (GHz)	 & 8.4851      &  4.8351     & 8.4851       & 8.4851      & 8.4851      & 8.4851      & 4.8351        & 4.8351         \\
Bandwidth (MHz)  &  50         & 50          &  50          &  50         & 50          &  50         &     50        &  50            \\
Flux Calibrator  &0521+166     & 1331+305    & 0521+166     & 1331+305    & 1331+305    & 1331+305    & 0137+331      &0137+331        \\
                 &             & 0521+166    &              &             & 0521+166    & 0521+166    &               &                \\   
Phase Calibrator &0244+624     & 0614+607    & 0244+624     & 0614+607    & 0244+624    & 0244+624    & 0614+607      & 0614+607       \\
                 &             &             &              &             & 0614+607    & 0614+607    & 0217+738      & 0217+738       \\
Instrumental     &0244+624     & 0319+415    & 0244+624     & 0614+607    & 0244+624    & 0244+624    & 0614+607      & 0614+607       \\
Polarization     &             & 0614+607    &              &             & 0614+607    & 0614+607    & 0217+738      & 0217+738       \\
Absolute         & 0521+166    & 1331+305    &0521+166      & 1331+305    & 1331+305    & 0521+166     & 0521+166     & 1331+305       \\
Polarization Angle &           &             &              &             &             &              &              &                \\
R-L Angle        & -20$^\circ$ & 66$^\circ$  & -20$^\circ$  &  66$^\circ$ &66$^\circ$   & -22$^\circ$  & -24$^\circ$  & 66$^\circ$     \\
\enddata
\end{deluxetable}

\begin{deluxetable}{ccccccc}
\tablewidth{0pt}
\tabletypesize{\scriptsize}
\tablecaption{Summary of Effelsberg 100-m Observations \label{sec:effelsberg_obs_summary} }
\tablehead{
  \colhead{Frequency} &
  \colhead{Bandwidth} &
  \colhead{} &
  \colhead{Beam} &
  \colhead{Map Size} &
  \colhead{Observing Time} &
  \colhead{Noise} \\
  \colhead{GHz} &
  \colhead{MHz} &
  \colhead{Number of Feeds} &
  \colhead{$\arcmin$} & 
  \colhead{$\arcmin$} &
  \colhead{hr} &
  \colhead{$\rm{mJy} \ \rm{beam}^{-1}$} }
\startdata
4.85   &  500   & 2  &  $2.7 \times 2.5$ & $10 \times 10$ & 5    & 2.5 \\
10.45  & 300    & 4  &  $1.1 \times 1.1$ & $12 \times 8$  & 26   & 0.5 \\
\enddata
\end{deluxetable}

\begin{deluxetable}{lccccc}
\tablewidth{0pt}
\tabletypesize{\scriptsize}
\tablecaption{Parameters of Previous Observations \label{tab:previous_obs}}
\tablehead{
\colhead{} &
\colhead{Wavelength} &
\colhead{Frequency} &
\colhead{Resolution} &
\colhead{Largest Angular Scale} &
\colhead{Sensitivity} \\
\colhead{Paper} &
\colhead{cm} &
\colhead{GHz} & 
\colhead{\arcsec} &
\colhead{\arcmin} &
\colhead{\mJybeam} 
}
\startdata
Seaquist \& Bignell (1976)  & 11.1 &  2.7 & $7.7 \times 5.5$ & 4   & 0.71 \mJybeam \\
                            & 3.7  &  8.1 & $3.0 \times 2.1$ & 1.3 & 0.23 \mJybeam \\ 
\hline
Condon (1983)               & 21.4 &  1.4 & $\sim 15-20$               & 15  & 0.04 \mJybeam \\ 
                            & 21.4 &  1.4 & $5.9 \times 5.8$ & 2   & 0.06 \mJybeam \\
\hline
Klein \& Gr{\"a}ve (1986)   & 1.2  & 24.5 & 29.7            & \nodata\tablenotemark{a} & 1.7 \mJybeam \\
\hline
Israel \& de Bruyn (1988)   & 50   &  0.6 & $56 \times 62$   & 48  & \nodata\tablenotemark{b} \\
                            & 21.4 &  1.4 & $24 \times 27$   & 14  & 0.65 \mJybeam \\
                            & 6    &  5.0 & $6.9 \times 7.6$ & 6   & \nodata\tablenotemark{b} \\
\hline
Lisenfeld et al. (2004)    & 20    & 1.5 & 6                & 2   & 0.12 \mJybeam \\
                           & 6.1   & 4.9 & 6                & 2   & 0.13 \mJybeam \\
                           & 3.6   & 8.4 & 6                & 2   & 0.09 \mJybeam \\
                           & 1.9   & 15.4 & 6               & 2   & 0.15 \mJybeam \\
\enddata
\tablenotetext{a}{ These are single-dish observations, which are sensitive to all spatial scales greater than their resolution.}
\tablenotetext{b}{ Sensitivities not given.}
\end{deluxetable}

\begin{deluxetable}{lcccc}
\tablewidth{0pt}
\tabletypesize{\scriptsize}
\tablecaption{Pressures of Various Components of the ISM of NGC 1569 \label{tab:ism_pressures}}
\tablehead{
\colhead{} &
\colhead{} &
\colhead{} &
\colhead{} &
\colhead{Pressure} \\ 
\colhead{Component} &
\colhead{Input Values} & 
\colhead{Equation} &
\colhead{Reference} & 
\colhead{$10^5\ \rm{K} \ \rm{cm^{-3}}$} }
\startdata
Magnetic Field (central) & $B= 38 \ \rm{\mu G}$ & $B^2/(8 \pi k)$ & 1 & 4.2 \\
Magnetic Field (halo)    & $B= 12 \ \rm{\mu G}$ & $B^2/(8 \pi k)$ & 1 & 0.41 \\
Hot Gas (central)        & $n_e = 0.049 \ \rm{cm^{-3}}$, $T=7.23\times10^6 \ \rm{K}$ & $2 n_e T$  & 2  & 7.2 \\
Hot Gas (halo)           & $n_e = 0.028 \ \rm{cm^{-3}}$, $T=7.23\times10^6 \ \rm{K}$ & $2 n_e T$  & 2  & 2.0 \\ 
HII Regions              & $n_e \sim 57 \ \rm{cm^{-3}}$, $T=1.0\times10^4 \ \rm{K}$ & $2 n_e T$   & 3  &  11 \\ 
Diffuse Ionized Gas (DIG) & equilibrium between hot gas and DIG  & --    & 4  &  1--6.3 \\
Warm Gas (turbulent, broad)     & $n_e \sim 0.1 \ \rm{cm^{-3}}$, $v_{turb} = 150 \ \rm{km \ s^{-1}}$ &  $n_e m_H v_{turb}^2 / (2 k)$ & 5 & 1.4 \\
Warm Gas (turbulent, narrow)        & $n_e \sim 57 \ \rm{cm^{-3}}$, $v_{turb} = 70 \ \rm{km \ s^{-1}}$ &  $n_e m_H v_{turb}^2 / (2 k)$ & 3, 5 & 170  \\
Gravity   & $v_{rad} = 65.5 \ \rm{km \ s^{-1}}$, $i=63\arcdeg$, $r=1.4 \ \rm{kpc}$ & $3 (v_{rad}/ \sin{i})^4 / (20 \pi r^2 G k)$ & 6 & 8.1 \\
\enddata
\tablerefs{ (1): This work, (2): \citet{ch2:2005MNRAS.358.1453O}, (3): \citet{ch2:1997ApJ...491..561M}, (4): \citet{1999ApJ...513..156M},
(5): \citet{ch2:2008MNRAS.383..864W}, (6): \citet{ch2:2005AJ....130..524M} }
\end{deluxetable}

\begin{deluxetable}{cllll}
\tablewidth{0pt}
\tabletypesize{\scriptsize}
\tablecaption{Magnetic Field Strengths in Starburst
  Galaxies \label{tab:bfield_strengths_starbursts}}
\tablehead{
  \colhead{} &
  \colhead{$B_t$} &
  \colhead{$B_{t,center}$} &
  \colhead{$B_{t,halo}$} &
  \colhead{} \\
  \colhead{Galaxy} &
  \colhead{\uG} &
  \colhead{\uG} &
  \colhead{\uG} &
  \colhead{Reference} \\
}
\startdata
M82      & 50      & \nodata       & \nodata       &  1  \\ 
NGC 253  & \nodata & 17            & 7             &  2 \\ 
NGC 4666 & \nodata & 14.4          & 7.1           &  3 \\ 
NGC 1808 & \nodata & 75            & 10--35        &  4 \\ 
NGC 1569 & \nodata  & 38           & 10-15         &  5 \\
\enddata
\tablenotetext{a}{These are estimates of the magnetic field strength based on equipartition between gravitational and magnetic pressures from \citet{2006ApJ...645..186T}.}
\tablerefs{ (1): \citet{ch2:1988A&A...190...41K}, 
  (2): \citet{ch2:1994A&A...292..409B}, (3): \citet{ch2:1997A&A...320..731D},
  (4): \citet{ch2:1990A&A...240..237D}, (5): This work   }
\end{deluxetable}


\clearpage

\end{document}